\documentclass[journal=apchd5, layout=twocolumn, manuscript=article]{achemso}

\usepackage[T1]{fontenc}
\usepackage{amssymb}
\usepackage[unicode=true, colorlinks=false]{hyperref} 
\usepackage{physics}
\usepackage{tikz} 
\usepackage{natbib} 

\usepackage{mhchem}
\usepackage{array}
\usepackage{booktabs}
\usepackage{listings}
\usepackage{lmodern}
\usepackage{microtype}
\usepackage{caption}
\usepackage{float}
\usepackage{geometry}
\usepackage{setspace}
\usepackage{xkeyval}
\usepackage{ulem}

\usepackage{pdflscape}
\usepackage{amsmath}
\usepackage{amssymb}
\usepackage{mathtools}
\usepackage{harpoon}
\usepackage[
  separate-uncertainty=true,   
  per-mode=symbol-or-fraction, 
]{siunitx}
\usepackage{xfrac}
\newcommand\sref[2]{\hyperref[#1]{\ref*{#1}(#2)}}
\newcommand\cunderline[2]{\textcolor{#1}{\underline{\textcolor{black}{#2}}}}
\usepackage{bibentry}

\newcommand{\be}{\begin{equation}}
\newcommand{\ee}{\end{equation}}
\newcommand{\bee}{\begin{eqnarray}}
\newcommand{\eee}{\end{eqnarray}}

\author{A.~V.~Trifonov}
\email{artur.trifonov@tu-dortmund.de}
\affiliation{Experimentelle Physik 2, Technische Universit\"at Dortmund, 44221 Dortmund, Germany}
\alsoaffiliation{Spin Optics Laboratory, St. Petersburg State University, 198504 St. Petersburg, Russia}
\author{S.~Grisard}
\affiliation{Experimentelle Physik 2, Technische Universit\"at Dortmund, 44221 Dortmund, Germany}
\author{A.~N.~Kosarev}
\affiliation{Experimentelle Physik 2, Technische Universit\"at Dortmund, 44221 Dortmund, Germany}
\alsoaffiliation{Ioffe Institute, Russian Academy of Science, 194021 St. Petersburg, Russia}
\author{I.~A.~Akimov}
\affiliation{Experimentelle Physik 2, Technische Universit\"at Dortmund, 44221 Dortmund, Germany}
\alsoaffiliation{Ioffe Institute, Russian Academy of Science, 194021 St. Petersburg, Russia}
\author{D.~R.~Yakovlev}
\affiliation{Experimentelle Physik 2, Technische Universit\"at Dortmund, 44221 Dortmund, Germany}
\alsoaffiliation{Ioffe Institute, Russian Academy of Science, 194021 St. Petersburg, Russia}
\author{J. H\"ocker}
\affiliation{Experimental Physics VI, Julius-Maximilian University of W\"urzburg, 
97074 W\"{u}rzburg, Germany}
\author{V. Dyakonov}
\affiliation{Experimental Physics VI, Julius-Maximilian University of W\"urzburg, 
97074 W\"{u}rzburg, Germany}
\author{M.~Bayer}
\affiliation{Experimentelle Physik 2, Technische Universit\"at Dortmund, 44221 Dortmund, Germany}
\alsoaffiliation{Ioffe Institute, Russian Academy of Science, 194021 St. Petersburg, Russia}

\title{Photon echo polarimetry of excitons and biexcitons in a CH$_3$NH$_3$PbI$_3$ perovskite single crystal}

\begin{document}

\begin{abstract}
Lead halide perovskites show remarkable performance when used in photovoltaic and optoelectronic devices. However, the peculiarities of light-matter interactions in these materials in general are far from being fully explored experimentally and theoretically. Here we specifically address the  energy level order of optical transitions and demonstrate photon echos in a methylammonium lead triiodide  single crystal, thereby determining the optical coherence times $T_2$ for excitons and biexcitons at cryogenic temperature to be 0.79 ps and 0.67 ps, respectively. Most importantly, we have developed an experimental photon-echo polarimetry method that not only identifies the contributions from exciton and biexciton complexes, but also allows accurate determination of the biexciton binding energy of 2.4 meV, even though the period of quantum beats between excitons and biexcitons is much longer than the coherence times of the resonances. Our experimental and theoretical analysis methods contribute to the understanding of the complex mechanism of quasiparticle interactions at moderate pump density and show that even in high-quality perovskite crystals and at very low temperatures, inhomogeneous broadening of excitonic transitions due to local crystal potential fluctuations is a source of optical dephasing.
\end{abstract}


\section*{Introduction}


Recently, the exceptional characteristics of lead halide perovskite materials essential for photovoltaics~\cite{Kojima-JAmChS2009,Jena-ChemRev2019,Jeong-Nat2021}, optoelectronics applications~\cite{Docampo-ACR2016,Murali-ACSMatLet2020,Fu-NatRevMat2019}, lasers~\cite{Eperon-EnEnvSci2014,Wang-NatCom2019,Wei-NatCom2019} and X-ray and gamma detectors~\cite{Nazarenko-NPGAsMat2017} have attracted the attention of a wide audience. 
Low temperature, solution and vacuum processable structures based on perovskite semiconductors are an attractive enrichment to conventional inorganic semiconductors. Owing to the impressive development of nanocrystal, film and crystal growth techniques~\cite{Zhang-ChemComm2016,Dey-ACSNano2021,Dong-Sci2015,Saidaminov-NatComm2015,Andrievich-AdvSci2021}, device performance has also reached a remarkable level. 
In photovoltaics, in only a few years, the power conversion efficiency (PCE) has rapidly increased from an initial value of 3.8\% to almost 25\% on laboratory-scale~\cite{Green-PiP2021}. Moreover, the first light-emitting electrochemical cells have been developed, and light-emitting diodes (LEDs) already exhibit internal quantum efficiencies exceeding 20\% and tuneable light emission spectra~\cite{Lin-Nature2018}.
 On the other hand, the knowledge currently available about the excited states of electrons and excitons, their properties and particularly interactions is far from sufficient, and certainly let alone complete. First steps in this direction have been taken mainly in low-dimensional systems~\cite{Himtermayr-ACSn2018,Ashner-ACSel2019,LI-NanoPhot,Huang-JPCL2020,Zhang-ChemComm2016}. Their complex and controversially discussed exciton fine-structure promises the discovery of fascinating physics~\cite{Fu-NanoLett2017,Sercel-NanoLett2019,Baranowski-AEM2020,Liu-PRL2020,Zhang-ChemComm2016} and efficient light sources.



The exact picture of the excitonic structure of bulk hybrid lead-halide perovskites that underlies properties of low-dimensional systems is unclear. 
In methylammonium lead  triiodide (CH$_3$NH$_3$PbI$_3$/ MAPbI$_3$)  and other hybrid perovskites, despite 25 years of research history~\cite{Hirasawa-PhB1994}, consensus on the exact value of the exciton binding energy, $\varepsilon_\mathrm{B}=10\ldots15$~meV, was found only recently~\cite{Baranowski-AEM2020}.  
Also properties of MAPbI$_3$ crystals and properties of carriers in them have been studied: carrier recombination rates and spin dephasing rates~\cite{Chen-JPCL2018}, accelerated relaxation of carriers due to a decrease in their Coulomb screening~\cite{Himtermayr-ACSn2018} and effective dielectric constant~\cite{Even-JPCC2014,Umari-SciRep2014, Brivio-PRB2014}, the influence of the huge temperature expansion coefficient on various mechanical and optical processes~\cite{Singh-JPCL2016}, the influence of dynamic processes in the crystal lattice on optical properties~\cite{Panzer-AEM2017}, a low rate of surface carrier recombination~\cite{Yang-NatComm2015}, phonon bottleneck phenomena at high optical excitation powers~\cite{Yang-NatPhot2016}.

Information on structure of energy transitions and coherent optical properties of hybrid lead-halide perovskites is barely known for bulk crystals and mostly based on studies of nanoplatelets~\cite{Bohn-ACSph2018} and thin films~\cite{March-SciRep2016, March-ACSPhot2017, March-JCP2019,Camargo-JACS2020,Webber-APL2017}. Exciton fine structure splitting of $\sim 200$ $\mu$eV has been measured in MAPbBr3 bulk crystal~\cite{Baranowski-NanoLett2019}. It was shown that organic-inorganic perovskites have interesting features associated with the organic cations.  The organic cation has rotational degrees of freedom and a dipole moment. Its random orientation and long range correlations of orientation lead to long-range potential fluctuations unlike in alloys or other conventional disordered systems~\cite{Ma-NanoLett2015}.  At high temperatures, this disorder is dynamic. At low temperatures, such disorder is associated with the frozen random fluctuations~\cite{March-JCP2019, March-ACSPhot2017}.  
The coexistence of domains with different crystallographic phases in a wide temperature range also leads to fluctuations in the band gap~\cite{Kong-RSOC2015,Whitfield-SciRep2016}.
These properties give rise to an inhomogeneous broadening even in high-quality crystals.

Coherent optical spectroscopy methods allow one to overcome inhomogeneous broadening.  For example, time-integrated four-wave mixing (TIFWM) gives $\varepsilon_\mathrm{B}=13$~meV in MAPbI$ _3 $~\cite{March-SciRep2016} hidden by inhomogeneous broadening, which is in good agreement with the conclusions of the review article~\cite{Baranowski-AEM2020}. Using TIFWM and coherent multidimensional spectroscopy, subbandgap defect-bound exciton states were discovered ~\cite{March-SciRep2016, Camargo-JACS2020}. In Ref.~\cite{March-ACSPhot2017} the TIFWM showed a weak interaction between excitons, as well as a long dephasing time of carriers. Carrier diffusion in MAPbI$ _3$ thin films was studied using four-wave mixing (FWM)~\cite{Webber-APL2017}.


In this work we use a variety of techniques of time- and spectrally-resolved FWM spectroscopy to study a bulk MAPbI$_3$ single crystal at low temperature $T=2$~K which shows photon echoes (PE) from excitons. Moreover, we have developed a transient photon echo polarimetry technique that allows us to unambiguously identify the biexciton resonance. The technique is based on the study of PE polarization oscillations induced by exciton-biexciton quantum beats. This approach enables us to determine the biexciton binding energy, $\varepsilon_\mathrm{XX}$, even if the quantum beats period exceeds the optical coherence times in the system under study and no complete amplitude oscillation can be observed. The PE polarimetry allows us to identify the exciton resonance with the optical coherence time $T_2^\mathrm{X} = 0.79$~ps, the  biexciton resonance with the coherence time $T_2^\mathrm{XX}=0.67$~ps and evaluate $\varepsilon_\mathrm{XX}=2.4$~meV in the MAPbI$_3$ crystal.

\begin{figure}
\includegraphics[width=\linewidth]{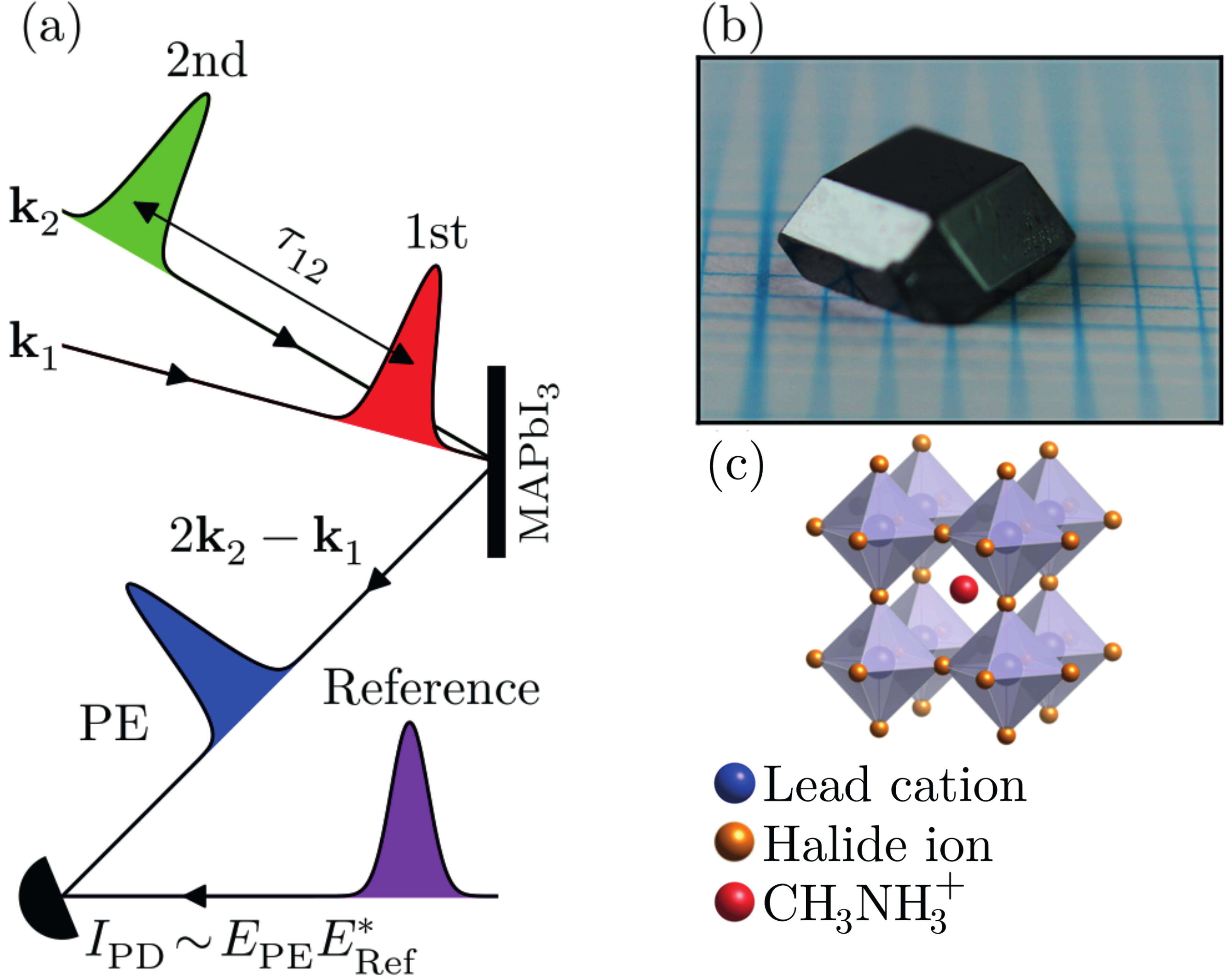}
\caption{ (a) Schematics of photon echo experiment and heterodyne detection. (b) Photo of the studied MAPbI$_4$ crystal on a mm-scale paper. (c) Crystallographic structure of the MAPbI$_3$ (taken from~\cite{phdthesisHocker}). 
}
\label{Fig0}
\end{figure}



\section*{Experimental results}

Degenerate transient FWM is a powerful experimental technique to gain insight into the nature and characteristics of optical excitations in semiconductors. Figure~\ref{Fig0}(a) shows schematically the experimental geometry while Figures~\ref{Fig0}(b,c) show the photo of the studied MAPbI$_3$ crystal and its crystallographic structure. The system under study is resonantly excited by a sequence of two optical pulses with wavevectors $\mathbf{k}_1$ and $\mathbf{k}_2$ delayed by the time $\tau_{12}$ with respect to each other. The third order susceptibility $\chi^{(3)}$ gives rise to a coherent response of the system in the phase matched direction $2\mathbf{k}_2-\mathbf{k}_1$. Note, $\chi^{(3)}$ has maxima around resonance frequencies as well as susceptibilities of other orders $\chi^{(1)}$, $\chi^{(2)}$, etc.~\cite{Shah-bookUltrafast} In our experiment this response is measured using heterodyne detection~\cite{Langer-PRL2012,Poltavtsev-PRB2016,Poltavtsev-PSS2018} where the recorded signal, $I_{\mathrm{PD}}$, is given by the interference between the measured light and a strong reference beam on the photodiode. By changing the reference pulse arrival time, $\tau_{\rm{ref}}$, we measure the dynamics of the FWM signals.

\begin{figure*}
\includegraphics[scale=1]{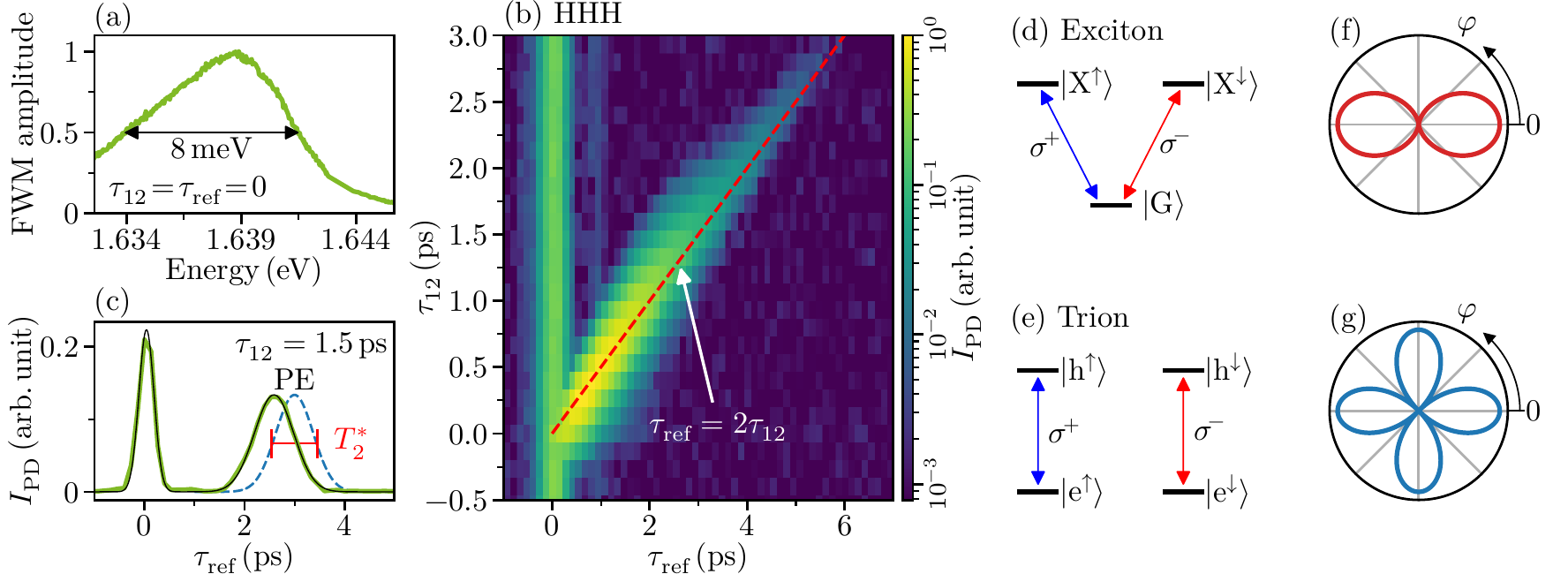}
\caption{ (a) FWM spectrum of MAPbI$_3$ crystal  measured at $T=2$~K. (b) The heterodyne signal (colour) as a function of  $\tau_{12}$ and $\tau_{\rm{ref}}$. Red dashed line tracks $\tau_{\rm{ref}} = 2\tau_{12}$. 170~fs laser pulses tuned to 1.638~eV covering the whole FMW spectrum shown in panel (a) are used. (c) Experimental $\tau_{\rm{ref}}$ dependence of the heterodyne signal for $\tau_{12}=1.5$~ps (green) and its fit by Gaussians (black). Blue dashed line is the PE pulse deconvoluted from the measured signal by taking into account the impact of short coherence time and the duration of the reference pulse.  (d)  Schematics of exciton energy levels, where $\mathrm{\ket{G}}$ is the ground state of unexcited crystal, $\mathrm{\ket{X^\uparrow}}$ and $\mathrm{\ket{X^\downarrow}}$ are the exciton states with spin up and spin down, respectively. (e) Schematics of negatively charged exciton (trion or donor bound exciton) energy levels, where $\mathrm{\ket{e^\uparrow}}$ and $\mathrm{\ket{e^\downarrow}}$ are ground states with resident electron spin up and spin down, $\mathrm{\ket{h^\uparrow}}$ and $\mathrm{\ket{h^\downarrow}}$ are the trion states with hole spin up and spin down.  Arrows  in (d) and (e) show allowed optical transitions in circular  polarizations $\sigma^{+}$ and $\sigma^{-}$. Modeled two-leaves (f) and four-leaves (g) polar rosettes corresponding to the exciton and trion (or donor bound exciton) level schemes in (d) and (e), respectively. 
}
\label{Fig1}
\end{figure*}

Figure~\ref{Fig1}(a) shows the spectral dependence of the FWM signal measured by scanning the wavelength of the spectrally narrow ($\sim0.3$~meV width) laser pulses with a duration of about 2.8~ps hitting the sample simultaneously ($\tau_{12}=\tau_{\rm{ref}}=0$). The spectrum shows a broad peak with full width at half maximum (FWHM) of 8~meV and maximum at 1.639~eV. These values are in good agreement with results of previous studies of excitons in MAPbI$_3$~\cite{Bohn-ACSph2018,Baranowski-AEM2020,March-ACSPhot2017}. All other experimental results presented below were obtained using a $\sim11$~meV spectrally broad laser (pulse duration 170~fs) with the spectrum covering the whole band shown in Figure~\ref{Fig1}(a). The central photon energy of the pulses was tuned to 1.638~eV. These 170~fs short optical pulses improve the temporal resolution of the experiment, while the 2.8~ps pulses provide better spectral resolution.

An important degree of freedom in our experiments is the polarization configuration. Hereafter, we use the two- or three-letter notation like HH and HHH for the polarization configuration, in which the first two letters describe the polarization of the first and second pulses. H and V correspond to horizontal and vertical linear polarizations, respectively. D and A are linear polarizations in the basis rotated by 45$^\circ$ with respect to H and V. $\sigma^+$ and $\sigma^-$ mark circular polarizations. The third letter corresponds to the detection polarization given by the polarization of the reference beam.

Figure~\ref{Fig1}(b) shows the dependence of the FWM signal on $\tau_{12}$ and $\tau_{\rm{ref}}$ in the HHH polarization configuration.  The vertical line at $\tau_{\rm{ref}}=0$  is an artefact signal arising from the cross-correlation of the scattered first pulse with the reference pulse. It shows that our time resolution and accuracy of the used technique is limited by the laser pulse temporal width. The FWM signal shifts with increasing $\tau_{12}$ marked with the arrow and the dashed line corresponding to $\tau_{\rm{ref}}=2\tau_ {12}$.  As one can see, the maximum of the signal follows the dashed line proving that the observed signal is a PE signal.   The cross-section of the FWM signal by the dashed line is a transient PE amplitude, which decay with an optical coherence time $T_2$ ($\approx 0.79$~ps as discussed below).
Figure~\ref{Fig1}(c) shows a horizontal cross-section of Figure~\ref{Fig1}(b) for $\tau_{12}=1.5$~ps demonstrating the typical PE pulse time profile.
Here, the maximum of the PE pulse is slightly shifted from the expected position (3~ps) towards smaller $\tau_{\rm{ref}}$ since it is distorted by the short coherence time $T_2$  that is comparable with the PE pulse duration.  The dashed line in Figure~\ref{Fig1}(c) shows the PE pulse time profile deconvoluted from the measured signal by taking into account the impact of the short $T_2$ and the duration of the reference pulse (see details in Sect 3.2 of SI~). The FWHM of the deconvoluted dependence corresponds to the macroscopical polarization decay time $T_2^* = 0.88\pm0.12$~ps of an ensemble  with inhomogeneous broadening $\Gamma_{inh}= 8 \ln{(2)} \hbar / T_2^* = 4.1\pm0.6$~meV, which is larger than the homogeneous broadening $\Gamma_{hom} = 2 \hbar/T_2 = 1.67$~meV. Our result directly shows that at low temperature, the inhomogeneous broadening exceeds the homogeneous broadening induced by scattering on characteristic phonons, impurities or resident carriers remarkable for MAPbI$_3$ crystals.
To the best of our knowledge, this is the first direct experimental observation of a photon echo in lead halide perovskite bulk crystals. In previous FWM studies of lead halide perovskite thin films~\cite{Bohn-ACSph2018,March-JCP2019}, the PE presence was only assumed. 
 


As a next step we identify the origin of the exciton complexes contributing the PE signal. The recently developed photon echo polarimetry is a powerful technique to distinguish different exciton complexes~\cite{Poltavtsev-SciRep2019} (free excitons, donor-bound excitons, charged excitons), which is a highly non-trivial problem in optical spectroscopy. In this technique, the PE amplitude (FWM signal at $\tau_{\rm{ref}}=2\tau_{12}$) is measured as a function of the angle $\varphi$ between the linear polarizations of the two excitation pulses. We label this experimental protocol by the notation HRH, where R marks the linear polarization that is rotated. The PE arising from the hypothetical V-type level scheme shown in Figure~\ref{Fig1}(d) typical for excitons gives rise to a $|\cos{\varphi}|$ dependence as shown in Figure~\ref{Fig1}(f), which we call two-leaves polar rosette in the following. In contrast, the negatively charged exciton (trion) or donor bound exciton has the level structure shown in Figure~\ref{Fig1}(e) giving rise to a $|\cos{2\varphi}|$ dependence shown in Figure~\ref{Fig1}(g), which we call four-leaves polar rosette in the following. 

\begin{figure*}
\includegraphics[scale=1]{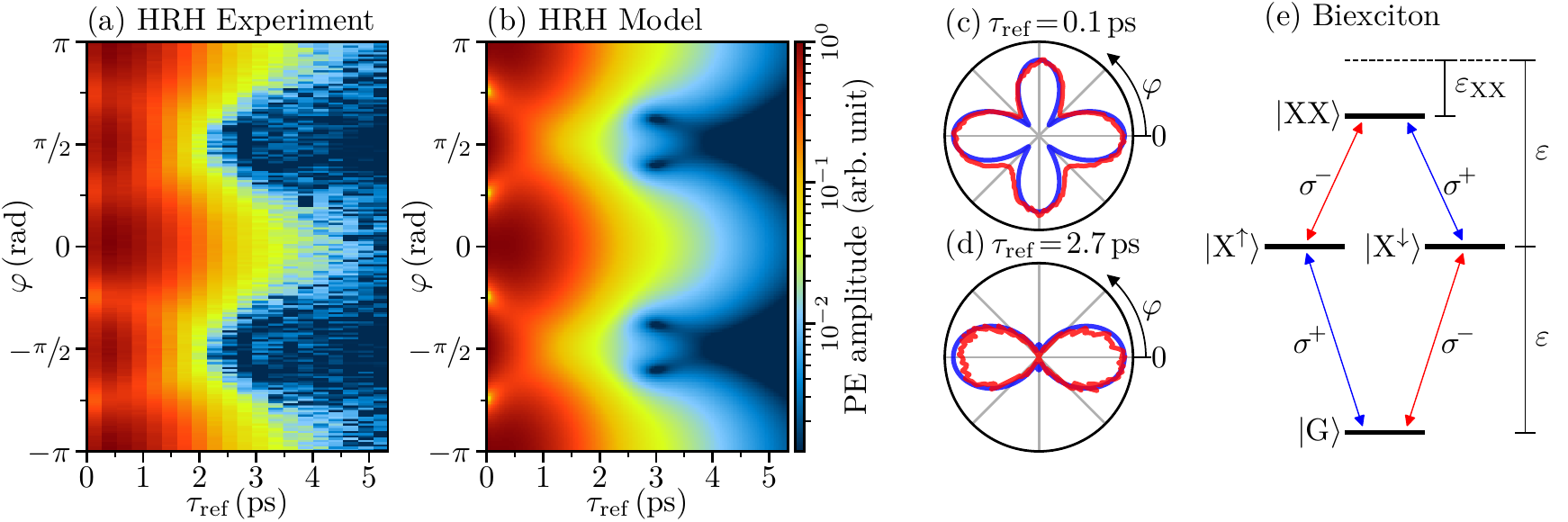}
\caption{ Experimental (a) and calculated (b) PE amplitude dependence on $\tau_{\rm{ref}}=2\tau_{12}$ and $\varphi$. (c,d) Polar rosettes measured (red) and calculated (blue) at $\tau_{\rm{ref}} = 0.1$~ps and $2.7$~ps. They are vertical cross-sections of the data in panels (a) and (b). Calculations are made in the biexciton model. (e) Diamond-like schematics of the exciton-biexciton energy levels, where $\mathrm{\ket{G}}$ is the ground state of the crystal, $\mathrm{\ket{X^\uparrow}}$ and $\mathrm{\ket{X^\downarrow}}$ are the excitons with opposite spin projections, $\mathrm{\ket{XX}}$ is the biexciton. Arrows show the allowed optical transitions in the circular polarizations.   
}
\label{Fig2}
\end{figure*}

In the studied MAPbI$_3$ crystal we found different types of rosettes at different $\tau_{\rm{12}}$ values. To study this effect in detail, we measure polar rosettes continuosly as a function of $\tau_{\rm{ref}} = 2\tau_{12}$. In this way, we expand the polarimetry technique by a further degree of freedom and arrive at two-dimensional data sets as visualized by the color map in Figure~\ref{Fig2}(a). As one can see, the four-leaves (peaks) behaviour is replaced by the two-leaves (peaks) exciton behaviour for $\tau_{\rm{ref}}>1$~ps. Figures~\ref{Fig2}(c,d) shows vertical cross-sections of Figure~\ref{Fig2}(a). For $\tau_{\rm{ref}} = 0.1$~ps the experimental polar rosette (red line in Figure~\ref{Fig2}(c)) resembles the theoretical four-leaves polar rosette (blue line), while for $\tau_{\rm{ref}} = 2.7$~ps an excitonic two-leaves polar rosette is observed, see Figure~\ref{Fig2}(d). {Based on the latest results of studying the resident carriers spin dynamics in lead halide perovskites~\cite{Belykh-NatCom2019,Kirstein-AdvMat2021} , one can assume the presence of trion or donor bound exciton transitions spectrally close to the free exciton resonance.} However, we show below that this behaviour is a signature of a exciton-biexciton system with a diamond-like level scheme shown in Figure~\ref{Fig2}(e). Here, the spin up $\mathrm{\ket{X^\uparrow}}$  and spin down $\mathrm{\ket{X^\downarrow}}$ exciton states are the initial states for optical excitation of the biexciton state $\mathrm{\ket{XX}}$.  The arrows denote the allowed optical transitions in the circular polarization basis, $\varepsilon$ is the exciton energy, and $\varepsilon_{\mathrm{XX}}$ is the biexciton binding energy, which is equal to the energy splitting of the exciton and biexciton in optical spectra. This system can be excited into a superposition exciton-biexciton state by the short optical pulse if FWHM of the laser pulses exceeds $\varepsilon_{\mathrm{XX}}$.

\begin{figure}
\includegraphics[scale=1]{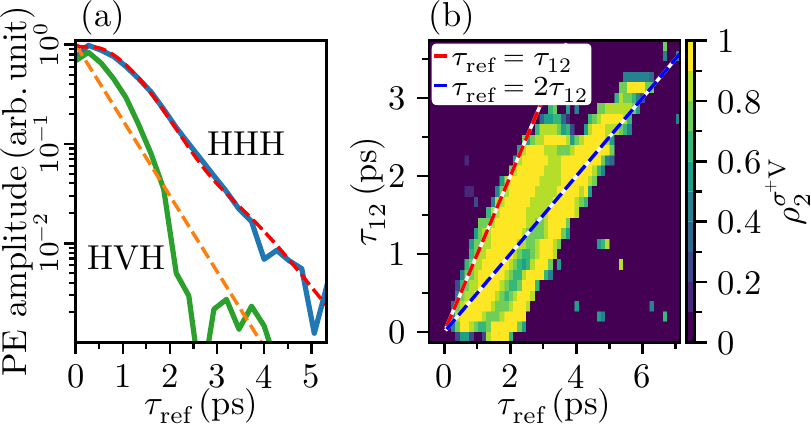}
\caption{ (a) Transient PE measured in HHH (blue) and HVH (green) polarization configurations that are horizontal cross-section of Figure~\ref{Fig2}(a) at $\varphi=0$ and $\pi/2$, respectively.  The transient PE calculated for the exciton-biexciton system in HHH (red dashed) and HVH (orange dashed) polarization configurations that are horizontal cross-sections of Figure~\ref{Fig2}(b) at $\varphi=0$ and $\pi/2$, respectively. (b) PE polarization contrast $\rho_2^{\sigma^+ V}$ as functions of $\tau_{12}$ and $\tau_{\rm{ref}}$. Red and blue dashed lines highlight the $\tau_{\rm{ref}} = \tau_{12}$ and $\tau_{\rm{ref}} = 2\tau_{12}$ dependences, respectively.
}
\label{Fig3}
\end{figure}

We have performed a theoretical analysis for exciton-biexciton system to predict its behaviour in transient PE polarimetry (see details in Sects. I and II of SI).  
The analytical expressions (Eqs. (S22,S23) in SI) for transient PE in the HHH and HVH polarization configurations
\bee
\label{PEHHH}
P_{\rm{HHH}}^\mathrm{XX} &\sim& [ 4 \exp(-\frac{2\tau_{\rm{ref}}}{T_2^X})+ \exp(-\frac{2\tau_{\rm{ref}}}{T_2^{XX}}) \\
&-&4\cos{\left( \frac{\omega_{QB} \tau_{\rm{ref}}}{2} \right)}\exp(-\frac{\tau_{\rm{ref}}}{T_2^{XX}}-\frac{\tau_{\rm{ref}}}{T_2^X}) ]^\frac{1}{2}. \nonumber
\eee
\be
 P_{\rm{HVH}}^\mathrm{XX} \sim \exp(-\frac{\tau_{\rm{ref}}}{T_2^{XX}})
 \label{PEHVH}
\ee
predict no oscillations in the HVH and damped oscillations in HHH polarization configuration. Here, $T^\mathrm{X}_{2}$ and $T_2^\mathrm{XX}$ are the exciton and biexciton coherence times, $\omega_\mathrm{QB} = \varepsilon_\mathrm{XX}/\hbar$ is the quantum beat frequency.  Figure~\ref{Fig3}(a) shows the experimental transient PE in the HHH (blue line) and HVH (green line) polarization configurations which are horizontal cross-sections of Figure~\ref{Fig2}(a) at $\varphi=0$ and $\pi/2$, respectively. As one can see, the transient PE amplitude in HHH configuration does not experience  oscillations with measurable amplitude in our experiment. The reason for this is the short coherence times of excitons and biexcitons. This assumption is confirmed by the simple estimate of the biexciton binding energy from analogy with hydrogen, for which the binding energy of the H$_2$ molecule is about $1/3$ of the H atom binding energy (4.7~eV/13.6~eV)~\cite{Hirschfelder-JOCP1950,Griessen-book}. 
 In Figure~\ref{Fig3}(a) the dashed lines are modeling results using Eqs.~\eqref{PEHHH},\eqref{PEHVH} with $T_2^X = 0.79\pm0.03$~ps and $T_2^{XX} = 0.67\pm0.03$~ps. These values were obtained from best fit of the HHH experimental data in Figure~\ref{Fig3}(a) with Eq.~\eqref{PEHHH}  and Figure~\ref{Fig4}(f) with Eq.~\eqref{NuVsT} (see details in Sect.~3.1 of SI). The noticeable non-exponential decay of the experimental HVH signal is most probably associated with the inhomogeneity of $\varepsilon_\mathrm{XX}$ and $T_2^{XX}$ due to crystal heterogeneity, which is not accounted in our model. The obtained exciton and biexciton coherence times are by almost 1.5 times shorter than the assumed quantum beat period $T_{QB} = h/\varepsilon_\mathrm{XX} = 3 h/ \varepsilon_\mathrm{B} = 1.1 $~ps.

The transient PE polar rosettes in the HRH configuration without decay is described by (Eq. (S20) in SI)
\be
P^\mathrm{XX}_{\rm{HRH}} = \sqrt{8 \cos^2{(\varphi)} \sin^2{\left ( \omega_\mathrm{QB} \tau_{\rm{ref}}/4 \right )} + \cos^2{(2\varphi)}}.
\ee
The second term on the right side yields a four-leaves polar rosette, while the first term of greater amplitude corresponds to the two-leaves polar rosette. It oscillates in time giving rise to the transient conversion between the four-leave polar rosette for $\omega_\mathrm{QB} \tau_{\rm{ref}}/4 = \pi n$, and the two-leave polar rosette for $\omega_\mathrm{QB} \tau_{\rm{ref}}/4 = \pi/2 + \pi n$, where $n\ge0$ is an integer number.
Because linearly polarized light at an angle $\varphi$ can be represented as a superposition of circularly polarized components $\mathrm{e}^{-\mathrm{i}\varphi}\sigma^+ + \mathrm{e}^{\mathrm{i}\varphi}\sigma^-$, the angle
$\varphi$ defines the phase between different quantum paths~\cite{mukamel1999principles} allowing them to interfere constructively ($\varphi = 0$) or destructively ($\varphi = \pi/2$).
%
%
A full analytical expression taking into account the decays of the exciton and biexciton coherences is given by Eq.~(S21) in the SI. Figure~\ref{Fig2}(b) shows the PE amplitude dependence on $\varphi$ and $\tau_{\rm{ref}}$  calculated in this model, using the experimentally obtained values of $T_2^X = 0.79$~ps and $T_2^{XX} = 0.67$~ps. The corresponding theoretical polar rosettes for $\tau_{\rm{ref}}=0.1$~ps and $\tau_{\rm{ref}}=2.7$~ps are shown in Figures~\ref{Fig2}~(c,d) by the blue curves. Thus, the experimentally observed change of the polar rosettes shape shown in Figures~\ref{Fig2}(a,c,d) is well explained by the exciton-biexciton quantum beats. Trion modeling performed in supplement cannot explain polar plots in Figure~\ref{Fig2} (for details see Figures S10(l,p) in SI).



The exciton-biexciton quantum beats manifest themselves in the discussed PE experiment with linearly polarized pulses as amplitude oscillations in the HHH polarization configuration. Because of the very short coherence times, the experimental data presented above do not allow us to  {prove the presence of quantum beats and to} estimate the biexciton binding energy from the PE amplitude beats (Eqs. (1)-(3)). However, the PE is characterised also by the polarization state which is described by the polarization contrasts (Stokes parameters) $\rho_{i}=  [|P_{\rm{ L_i}}|^2 - |P_{\rm{ M_i }}|^2]/[|P_{\rm{ L_i}}|^2 + |P_{\rm{ M_i}}|^2] $, where the $\rm{L_i}$ and $\rm{M_i}$ are pairs H/V, D/A and $\sigma^+$/$\sigma^-$ of detection polarizations for $i=1,2$ and 3.
A detailed study of the PE polarization state dynamics with correctly chosen polarization configuration of the incident pulses can provide additional information~\cite{Bolger-PRB1996,Paul-JOSAB1996,Tookey-JOSAB1998,Smirl-OptLett1998,Aoki-PRB2001}. As we show below, the PE in the polarization configuration with circular and linear polarization of the incident pulses experiences polarization state beats that are independent on the PE amplitude decay.  Figure~\ref{Fig3}(b) shows the experimental $\tau_{12}$ and $\tau_{\rm{ref}}$ dependence of the most informative (explained below) FWM polarization contrast  $\rho_2^{\rm{\sigma^+ V}} = (|P_{\rm{\sigma^+ VD}}|^2-|P_{\rm{\sigma^+ VA}}|^2)/(|P_{\rm{\sigma^+ VD}}|^2+|P_{\rm{\sigma^+ VA}}|^2)$.  Here, the polarization contrast values obtained with using a small FWM signal amplitude comparable with the noise level are set to zero (dark blue color). As one can see, the $\tau_{\rm{ref}}$ position of the central minimum of the FWM polarizations contrast oscillation (green area between two yellow areas) is parallel  to the $\tau_{\rm{ref}} = \tau_{12}$ (red dashed line) dependence. This behaviour points to the effect of quantum beats and excludes the possibility of interference of polarizations of two independent systems (e.g. exciton and trion) for which one can expect the dependence $\tau_{\rm{ref}} = 2\tau_{12}$  (blue dashed line)~\cite{Koch-PRL1992}. Section~4 of the SI provides additional arguments for this conclusion.

\section*{Transient FWM polarimetry of biexcitons}

We developed a transient polarimetry PE technique that allows us to reveal the biexciton optical resonance independently and measure its binding energy.  The technique is based on the detailed analysis of the transient polarization state of the PE that is modulated by excton biexciton quantum beats. We analyze the temporal behaviour and orientation of the polar rosettes in the $\sigma^{\pm}$RH polarization configuration with {{circularly}} polarized {{first}} pulse . 

\begin{figure}
\includegraphics[scale=1]{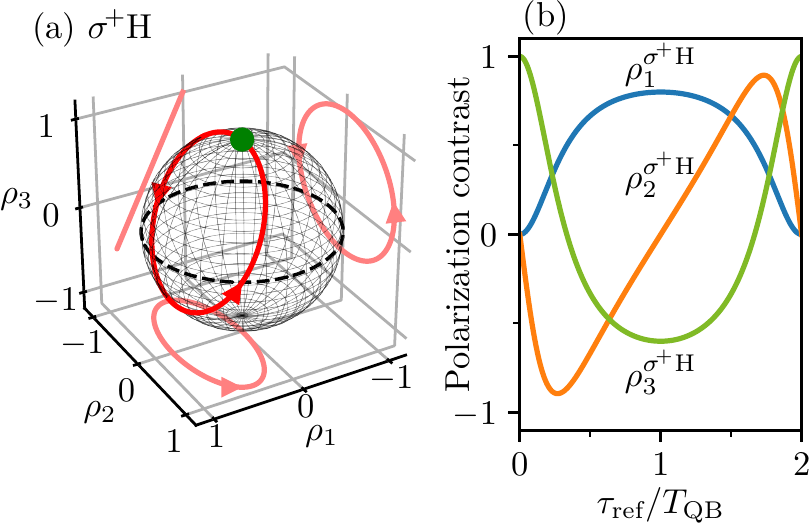}
\caption{ Time evolution of the PE polarization state for $\tau_{\rm{ref}}=2\tau_{12}$ in $\sigma^+$H polarization configuration modeled in the exciton-biexciton model without damping and shown in terms of the trajectory on the Poincaré sphere (a) and in terms of the polarization contrast parameters (b). {The light red curves on $\rho_1 \rho_2$, $\rho_1 \rho_3$ and $\rho_2 \rho_3$ planes in (a) are projections of the trajectory.}
}
\label{Fig5}
\end{figure}

To explain the technique's principle, we simulate the dynamics of the polarization state of the PE excited with a circularly polarized first pulse and a linearly polarized second pulse, namely for the $\sigma^+$H configuration. Figure~\ref{Fig5}(a) shows these dynamics on the Poincaré sphere modeled in the exciton-biexciton model without damping (see details in Sect.~2 of SI). Here, the quantum beats period corresponds to one rotation of the PE polarization state around the red circle. The starting and ending points (green dot) correspond to the $\sigma^+$ polarization of the first pulse. The precession direction and the trajectory orientation on the sphere is controlled  by the H polarization of the second pulse (see details in Sect.~2.3 of  SI). Figure~\ref{Fig5}(b) shows the dynamics  of the corresponding polarization contrasts $\rho_{i}^{\rm{\sigma^+ H}}$, where the superscript marks the polarizations of incoming pulses. Note, here one quantum beat period corresponds to $\tau_{\rm{ref}}/T_{QB}=2$ because $\tau_{\rm{ref}}/\tau_{12} = 2$ (maxima of the PE pulse). The dynamics of the PE elliptical polarization state in the range $0.3\lesssim\tau_{\rm{ref}} / (T_{QB}) \lesssim 1.7$ shows practically no change in $\rho_1^{\sigma^+ H}$ and $\rho_3^{\sigma^+ H}$, while $\rho_2^{\sigma^+ H}$ rises almost linearly from about $-1$ to $1$. The latter dependence highlights $\rho_2$ as the most informative polarization contrast parameter. One can describe the $\sigma^+$H PE polarization dynamics as rotation of the orientation of the main axis of the elliptically polarized PE from A to D through the H state.  In experiment, the main axis orientation can be measured as maximum of $\phi$ dependence, where $\phi$  is the angle of linear polarization of the detection. This experiment corresponds to the $P_{\rm{\sigma^+ H R}}$ dependence, where the symbol R marks the rotation of orientation of the linearly polarized detection.

The general equation describing the PE amplitude for the case of a $\sigma^{\pm}$ polarized first pulse, linearly polarized second pulse and {linearly polarized} detection is given by
\bee
\label{Psff}
P^\mathrm{XX}_{\sigma^{\pm} \varphi \phi} \sim [ 3 -2\cos{\left( \frac{\omega_\mathrm{QB}\tau_{\rm{ref}}}{2}\right)} - \\ 
-2 \sin{\left( 2\varphi + 2\phi \mp \frac{\omega_\mathrm{QB} \tau_{\rm{ref}}}{4} \right)} \sin{\left ( \frac{\omega_\mathrm{QB} \tau_{\rm{ref}}}{4} \right)} ]^{\frac{1}{2}} \nonumber.
\eee
Here $\varphi$ is the angle of linear polarization of the second pulse, and  $\phi$  is the angle of linear polarization of the detection. As one can see in Eq.~\eqref{Psff} the angles $\varphi$ and $\phi$ are equivalent. The $\varphi$ dependence for $\phi=0$ corresponds to the polar rosettes $\sigma^\pm$RH. Here, the rotation of the second pulse polarization gives rise to rotation of the whole trajectory on the Poincaré sphere about the $\rho_3$ axis detected in H polarization (see details in Sect.~2.3 of SI). Thus, the $\sigma^\pm$RH and $\sigma^\pm$HR dependencies for the exciton-biexciton system are equal, which we have confirmed  experimentally, see Figure~S3 in SI.

\begin{figure*}
\includegraphics[scale=1]{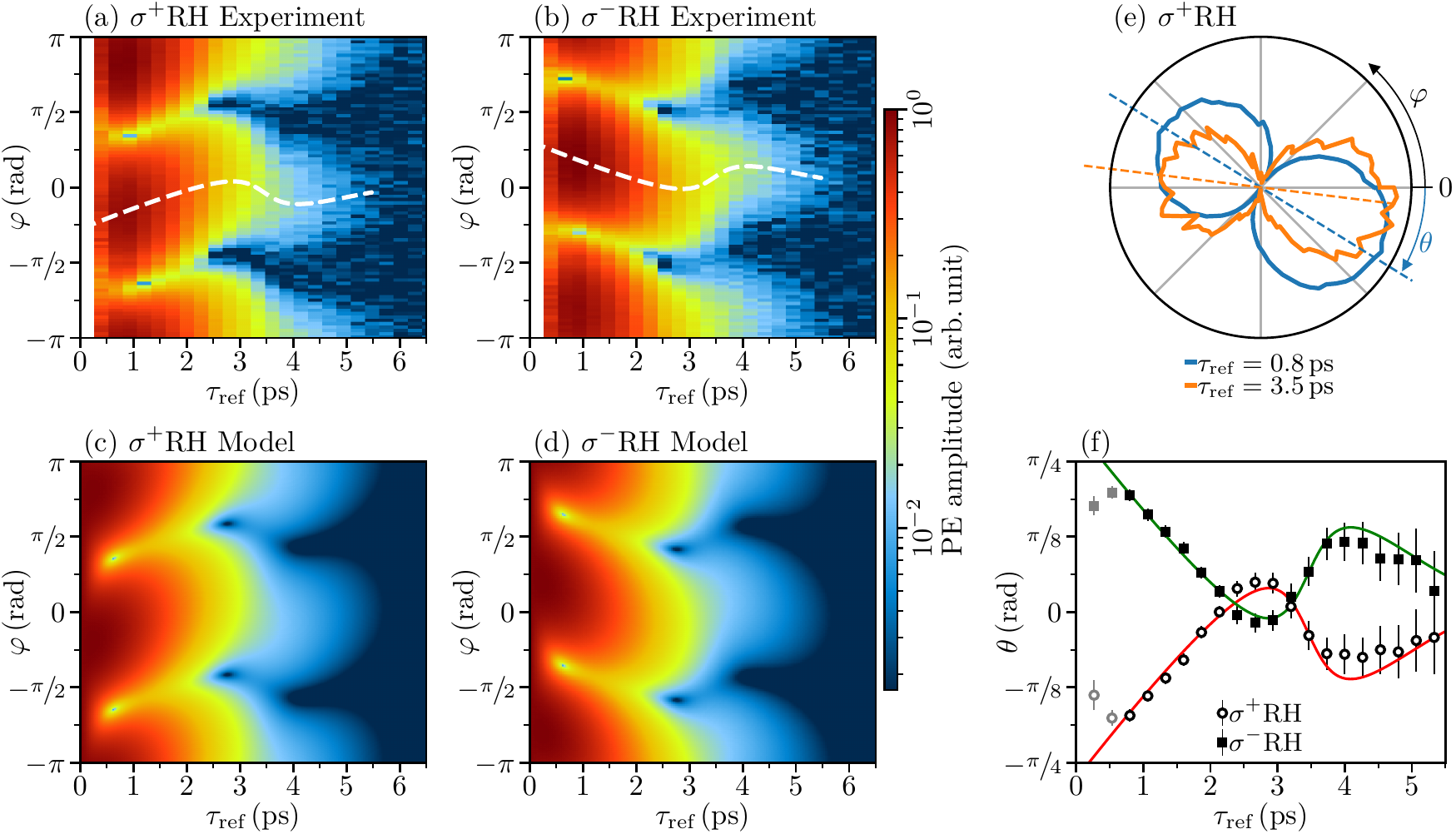}
\caption{ Experimental (a,b) and calculated in the exciton-biexciton model (c,d) transient polar rosettes in $\sigma ^+$RH and $\sigma^-$RH polarization configurations. (e)  $\sigma ^+$RH polar rosettes measured at $\tau_{\rm{ref}} = 0.5$~ps (blue) and $\tau_{\rm{ref}} = 3.5$~ps (orange).  (f) $\tau_{\rm{ref}}$ dependence of $\sigma ^+$RH and $\sigma^-$RH polar rosettes orientation angle $\theta$ obtained from experiment analysis (symbols) and their fitting with Eq.~(\ref{NuVsT}) (lines) with $T_2^\mathrm{X} = 0.79$~ps, $T_{2}^\mathrm{XX} = 0.67$~ps and $T_{QB} = 2\pi / \omega_\mathrm{QB} = 1.7$~ps.  
}
\label{Fig4}
\end{figure*}

In many cases, it is experimentally easier to vary the polarization of the second pulse than the detection polarization. Thereby the study of the polar rosettes behaviour in the $\sigma^{\pm}$RH polarization configuration is advantageous. 
Figures~\ref{Fig4}(a,b) shows such experimental dependences demonstrating a different behaviour for the $\sigma^+$ and $\sigma^-$ polarizations of the first pulse. { In contrast to the case of linearly polarized first pulse shown in Figures~\ref{Fig2}(a), these dependences have two peaks (leaves) because of circular polarization of the first pulse.} The $\tau_{\rm{ref}}$ position of peaks shifts with a constant rate in the range $0.5$~ps$\lesssim \tau_{\rm{ref}} \lesssim 2.5$~ps. In Figure~\ref{Fig4}(a,b), the shift is highlighted by a white dashed line marking the maximum of the signal amplitude. The shift corresponds to a rotation of the polar rosettes orientation. Figure~\ref{Fig4}(e) shows examples of polar rosettes $\sigma^+$RH at $\tau_{\rm{ref}} = 0.8$~ps and $3.5$~ps. Their orientation is marked by the dashed lines, while the orientation angle, $\theta$, is counted from the H axis.

To quantify the rotational behaviour of the $\sigma^{\pm}$RH rosettes we analyse the experimental data in Figures~\ref{Fig4}(a,b) using the fit function $|cos(\varphi-\theta)|$ for each $\tau_{\rm{ref}}$ value. The  circles and squares in Figure~\ref{Fig4}(f) show the $\tau_{\rm{ref}}$ oscillatory behaviour of $\theta$ for the $\sigma^+$RH and $\sigma^-$RH configurations, respectively. The time evolution is opposite for $\sigma^+$ and $\sigma^-$ polarized first pulse and the angle of initial orientation $\theta(\tau_{\rm{ref}}=0)\approx \mp \pi/4$ (A and D polarizations). Our calculations of the PE dependences on $\varphi$ and $\tau_{\rm{ref}}$ in $\sigma ^+$RH and $\sigma^-$RH polarization configurations for other level schemes (see details in Sects. 3 and 4 of the SI) show that the opposite behaviour for $\sigma^+$ and $\sigma^-$ polarized first pulse and the corresponding $\theta(0) = \mp \pi/4$ is a unique property of the exciton-biexciton system. Our additional modeling (see Fig. S11 in the SI ) shows that fine structure splitting of the exciton discussed in Ref.~\cite{Baranowski-NanoLett2019} has unnoticeable influence while it is much smaller of $\varepsilon_{\mathrm{XX}}$ and exciton homogeneous broadening $ \Gamma_{\mathrm{X}}= 2 \hbar/T_2^{\mathrm{X}}$.

The modeling of the $\varphi$ and $\tau_{\rm{ref}}$ dependence of the PE amplitude for the $\sigma^{\pm} $RH polarization configurations taking into account the exciton and biexciton decays gives the following analytical expression (Eq. (S32) in SI) for the time evolution of $\theta$ 
\be
\theta_{\rm{\sigma^{\pm}RH}} = \pm \frac{1}{2} \arctan{\left ( \frac{\sin{\left( \frac{\omega_\mathrm{QB} \tau_{\rm{ref}}}{2} \right)}}{\cos{\left( \frac{\omega_\mathrm{QB} \tau_{\rm{ref}}}{2} \right)} - \exp( \frac{\tau_{\rm{ref}}}{T_\mathrm{\Delta}})} \right)},
\label{NuVsT}
\ee
where $1/T_\mathrm{\Delta} = 1/T_2^\mathrm{XX} - 1/T_2^\mathrm{X}$.
The best fit of the experimental data in Figure~\ref{Fig4}(f) by Eq.~\eqref{NuVsT} (lines) and of the HHH decay in Figure~\ref{Fig3}(a) by Eq.~(\ref{PEHHH}) gives $T_2^\mathrm{X}=0.79\pm0.03$~ps, $T_2^\mathrm{XX}=0.67\pm0.03$~ps and $\varepsilon_\mathrm{XX} =2.4\pm0.2$~meV ($T_{QB} \approx 1.7$~ps). Note, for the fits we introduced a small shift of all dependencies by $\theta_0\approx0.1$ rad to take into account systematic errors associated with the finite pulse duration, as well as imperfections of the detection polarization and circular polarization of first pulse (see details in Sect.~3.1 of SI). We also excluded from the fit the first two experimental points shown by the grey symbols in Figure~\ref{Fig4}(f) because of the same reason.

\section*{Discussion and conclusions}

Figures~\ref{Fig4}(b,d) show the modeling in the exciton-biexciton model. The excellent agreement of all experimental data and the modeling allows us to conclude that the excitons and biexcitons are responsible for the PE signals in a bulk MAPbI$_3$ crystal. We evaluate their coherence time and the biexciton binding energy. To the best of our knowledge, our work is the first direct demonstration of biexcitons and measurement of their binding energy in MAPbI$_3$ crystals. The obtained $\varepsilon_\mathrm{XX}$ value of 2.4 meV is close to the rough estimate using the hydrogen molecule analogy, and the ratio $\varepsilon_\mathrm{XX}/ \varepsilon_\mathrm{B}\approx0.2$ is in good agreement with results of study of the biexcitons in the conventional bulk semiconductors~\cite{Klingshirn-PhysRep1981}. Note that free biexciton can be observed not in all bulk semiconductors~\cite{Klingshirn-PhysRep1981,Bressanini-PRA1998}. Probably, the localization of excitons, which is manifested in noticable ihomogeneous broadenig of optical transitions, simplifies the biexciton formation. Thereby the biexcitons are only weakly bound since the binding energy is comparble with homogeneous linewidth of biexciton-exciton optical transition.

The developed transient PE polarimetry technique is a powerful tool for biexciton identification and $\varepsilon_\mathrm{XX}$ measurement. It has several important advantages. First, the technique allows one to overcome the inhomogeneous broadening of the optical transitions in the system under study. Second, the PE polarization state measurements are free of the impact of the PE amplitude decay. Here, special interest is attracted by the case when the coherence times are shorter than the beat period. The oscillations of the $\sigma^{\pm}$RH polar rosettes orientation decay with rate the $1/T_\mathrm{\Delta} = 1/T_2^\mathrm{XX} - 1/T_2^\mathrm{X}$ (see Eq.~(\ref{NuVsT})), i.e. with the {\it{difference}} of coherence decay rates of biexciton and exciton. 
The oscillations of the PE amplitude decay with a rate equal to the {\it{sum}} of the decay rates $1/T_\mathrm{\Sigma}= 1 / T_2^\mathrm{XX} + 1 / T_2^\mathrm{X}$ (see Eq.~\eqref{PEHHH}). For equal coherence decay rates ($T_2^\mathrm{X} = T_2^\mathrm{XX}$) the oscillations of the polar rosettes orientation have no decay, which limits the technique only by the dynamic range of the PE amplitude measurements. In the spectral domain the properties of the technique looks more intriguing. $T_2^\mathrm{X}$ and $T_2^\mathrm{XX}$ give the homogeneous linewidth $\Gamma_\mathrm{X(XX)} = 2 \hbar/T_2^\mathrm{X(XX)}$ of the exciton and biexciton resonances split by $\varepsilon_\mathrm{XX}$. According to the above discussion the potential of the suggested technique to measure $\varepsilon_\mathrm{XX}$ is limited by $2\varepsilon_\mathrm{XX} > \Gamma_\mathrm{XX} - \Gamma_\mathrm{X}$. If $\Gamma_\mathrm{XX} = \Gamma_\mathrm{X}$ then exciton and biexciton lines can be always resolved by this technique. In the system under study $\Gamma_\mathrm{XX} =1.97 \pm 0.02$~meV and  $ \Gamma_\mathrm{X} = 1.67\pm0.02$~meV which gives the limit for the smallest measureable $\varepsilon_\mathrm{XX}$ as 0.15~meV. 
This property makes the technique very  helpful for studying new materials, e.g. with a strong inhomogeneous broadening, as well as short coherence times of the exciton complexes. For example, at high temperatures or in systems with a high concentration of defects, the coherence times are governed by scattering processes and are expected to be equal. 




\section*{Methods}
\textbf{MAPbI$_3$ crystals.}
MAPbI$_3$ single crystals were low temperature solution grown in a reactive inverse temperature crystallization (ITC) process~\cite{hocker2021}. As compared to the pure inverse temperature crystallization instead the pure $\gamma$-butyrolactone (GBL) precursor solvent, an alcohol-GBL mixture was used. The polarity of the mixed precursor changes, which leads to the lower solubility of the dissociated perovskite MAPbI$_3$, and thus to an optimization of the nucleation rate. Accordingly, the crystallization takes place at lower temperatures compared to the conventional ITC method. Black MAPbI$_3$ single crystals were obtained at a temperature of 85$^\circ$C instead of about 110$^\circ$C. At room temperature a tetragonal phase with lattice constant $a=0.893$~nm and $c=1.25$~nm was determined with X-ray diffraction (XRD) technique~\cite{hocker2021}. The size of the crystal is about $2\times 2 \times 2$~mm$^3$. The crystal shape is noncuboid, but the crystal structure exhibits arisotype cubic symmetry. The front facet was X-ray characterized to point towards the a-axis \cite{hocker2021}.

\textbf{Experimental details.} 
The sample was cooled down in a liquid helium bath cryostat to a temperature of 2~K.  A piezo-mechanical translator (attocube) allows us moving the sample to find surface areas with a mirror like reflection (typical size of 300 $\mu$m). All  optical pulses are generated by a Ti:Sapphire laser. They have either $\approx2.8$~ps or $\approx170$~fs duration and a repetition rate of 75.75~MHz. The time delays between the pulses are changed using mechanical delay lines. The experimental geometry is illustrated in Figure~\ref{Fig1}(a). The laser pulses were focused to a spot of about 200~$\mu$m diameter using an 0.5~m spherical metallic mirror. The power of the first beam is 1~mW, the power of the second beam is 0.8~mW. The incidence angles of the pulses are close to normal and equal to $\approx 1/50$~rad and $\approx 2/50$~rad (corresponding to the in-plane wavevectors $\mathbf{k}_1$ and $\mathbf{k}_2$).  The PE pulses were measured in reflectance geometry in the direction $\approx 3/50$~rad, which corresponds to the PE wavevector $2\mathbf{k}_2-\mathbf{k}_1$. Optical heterodyne detection was used to perform time-resolved PE experiments and to enhance the detected signals~\cite{Langer-PRL2012,Poltavtsev-PRB2016,Poltavtsev-PSS2018}. By mixing with a strong reference pulse (0.5~mW) and scanning the time delay between the first pulse and the reference pulse, $\tau_{\rm{ref}}$, one can measure the temporal profile of the photon echo pulse. The simultaneous scan of $\tau_{12}$ and $\tau_{\mathrm{ref}}=2\tau_{12}$ allows one to measure the decay of the PE amplitude. Here we use the two- or three-letter notations like HH or HHH for the polarization configuration, in which the first two letters correspond to the polarizations of the first and second pulses. H and V correspond to linear horizontal and vertical polarizations. D and A form the linear basis rotated by 45$^\circ$ with respect to H and V. $\sigma^+$ and $\sigma^-$ mark circular polarizations. The third letter corresponds to the detection polarization given by the polarization of the reference beam. Polar rosettes were measured using motorized stages rotating half-wave plates.

\textbf{Supplementary Information}
The detailed description of the theoretical model; results of analytical calculations for exciton-biexciton system; documentation of fitting procedure; additional modelling results.

\textbf{Notes} \\
The authors declare no competing financial interest.

\section*{Acknowledgement}
The authors are grateful to Natalia Kopteva and Erik Kirstein for discussing and sharing reflectance and photoluminescence spectra.
The authors acknowledge financial support by the Deutsche Forschungsgemeinschaft in the frame of the Priority Programme SPP 2196 (Project YA 65/26-1  and DY18/15-1) and the International Collaborative Research Centre TRR 160 (Project A3). It was also supported by the Russian Foundation for Basic Research (Project No. 19-52-12046). A.V.T. acknowledges the Saint Petersburg State University (Grant No. 73031758). J. H. and V.D. acknowledge financial support from the DFG through the Würzburg-Dresden Cluster of Excellence on Complexity and Topology in Quantum Matter—ct.qmat (EXC 2147, project-id 39085490) and from the Bavarian State Ministry of Education and Culture, Science and Arts within the Collaborative Research Network “Solar Technologies go Hybrid“. 




\clearpage

\onecolumn

\def\theequation{S\arabic{equation}}
\def\thefigure{S\arabic{figure}}
\def\thetable{S\arabic{table}}
\setcounter{equation}{0}
\setcounter{figure}{0}
\setcounter{table}{0}

\begin{center}
\Huge{\textbf{Supplementary information}}\end{center}
\doublespacing

\section{Modeling of photon echo}
For a better understanding of the transient polarization properties of the photon echo,
we use a perturbative model that allows us to calculate the photon echo amplitude 
of various exciton complexes as a function 
of time and the polarizations of all involved pulses. In this Section, 
we describe the modeling procedure in detail.

\subsection{General modeling procedure}\label{sec: modeling}
\begin{figure}
\centering
\includegraphics[scale = 1]{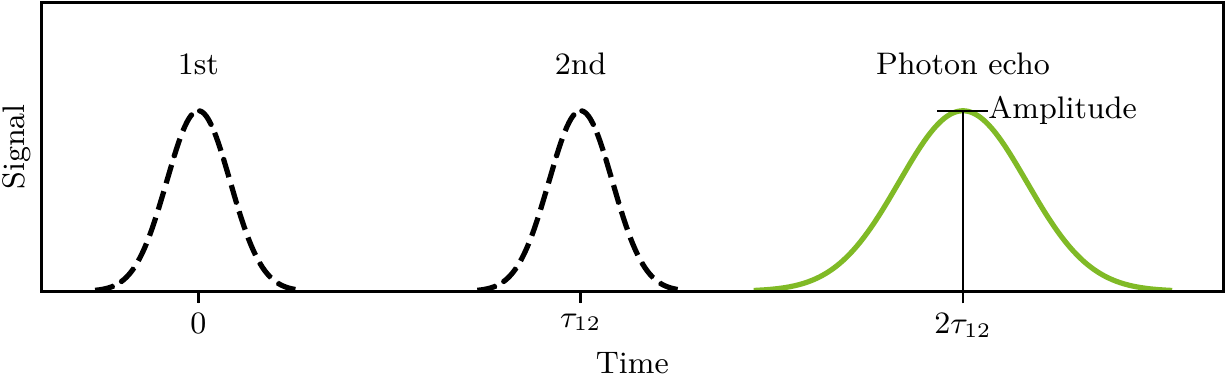}
\caption{Temporal arrangement of the optical pulses for a photon echo experiment.}
\label{fig: PE_scheme}
\end{figure}

Figure~\ref{fig: PE_scheme} shows the temporal arrangement of the optical pulses 
for a photon echo experiment. Two pulses 
impinge on the sample with a temporal delay of $\tau_{12}$ giving rise to 
the photon echo at time $2\tau_{12}$ after the first pulse. 
Our modeling procedure aims to calculate the 
amplitude of the photon echo pulse 
(as defined in Figure~\ref{fig: PE_scheme}) and, in particular, 
how it depends on the polarizations of the two incident pulses 
and $\tau_{12}$. For simplicity, we neglect the 
temporal shape and finite width $t_p$ of the pulses. In this 
way we also do not account for the finite temporal overlap between 
the pulses for small values of $\tau_{12} < t_p$.

The photon echo signal is the result of a macroscopic polarization $P$ of the sample, which is given 
by the expectation value of the dipole operator $\mathbf{d}$ 
\begin{equation}
    P = \text{Tr}\left( \mathbf{d} \boldsymbol{\rho} \right),
    \label{eq: expectation_value}
\end{equation} 
with the density matrix of the system $\boldsymbol{\rho}$. 
The density matrix behaves in time according to 
\begin{equation}
    i  \hbar \frac{\text{d}}{\text{d}t} \boldsymbol{\rho} = \left[\mathbf{H}, \, \boldsymbol{\rho}\right], 
\end{equation}
with the Hamilton operator $\mathbf{H}$, which itself can be written as the sum 
\begin{equation}
    \mathbf{H} = \mathbf{H}_0 + \mathbf{V}. 
\end{equation}
Here, $\mathbf{H}_0$ is the Hamilton operator of the unperturbed system, whereas $\mathbf{V}$ accounts for 
the light-matter interaction. 
All operators are represented by $N\!\times\!N$ matrices in the eigenbasis of the system, determined 
by $\mathbf{H}_0$ with eigenvalues 
$\varepsilon_1, \dots \varepsilon_N$.
To model the photon echo signal, we have to evaluate how $\boldsymbol{\rho}$ changes under the action of the first
and second pulses and how it evolves freely between the pulses. 
In particular, we arrive at an expression for the photon echo signal by pursuing the following steps: 
\begin{enumerate}

\item Let $\boldsymbol{\rho}^\text{b1}$ be the initial state of the crystal before arrival of the first laser pulse. 
We assume short pulse durations $t_p$ and all optical frequencies to be much higher 
than the eigenfrequencies of the system. Therefore, 
we can neglect the effect of $\mathbf{H}_0$ during the action of the laser pulses and calculate the density matrix 
after the action of the first pulse $\boldsymbol{\rho}^\text{a1}$ in second order as 
\begin{equation}
    \boldsymbol{\rho}^\text{a1} \approx \boldsymbol{\rho}^\text{b1} + \frac{i}{\hbar} \left[\boldsymbol{\rho}^\text{b1} + \frac{i}{\hbar} \left[\boldsymbol{\rho}^\text{b1}, \mathbf{V}_1\right] t_p, \mathbf{V}_1\right]t_p.
    \label{eq: rho^a1}
\end{equation}
We set $t_p \equiv 1$ in the following.

\item Between the pulses, we neglect the action of $\mathbf{V}$ on the system. The free evolution of the system can then 
be obtained by solving the equation 
\begin{equation}
    i \hbar \frac{\text{d}}{\text{d}t} \boldsymbol{\rho} = \left[\mathbf{H}_0, \, \boldsymbol{\rho}\right]. 
\end{equation}
Since $\mathbf{H}_0$ is diagonal, the entries of the density matrix before the second pulse, 
after time $\tau_{12} = \tau_\text{ref} / 2$ can be written in a closed form as
\begin{equation}
    \boldsymbol{\rho}^\text{b2}_\mathrm{ij} = \boldsymbol{\rho}^\text{a1}_\mathrm{ij} \exp\left\{ i \frac{\tau_{12}}{\hbar} \left[\mathbf{1}, \mathbf{H}_0 \right]_\mathrm{ij} \right\}, \quad \mathrm{i, j} = 1, \dots, N,  
    \label{eq: free_evolution}
\end{equation} 
with the matrix of ones $\mathbf{1}$ ($\mathbf{1}_\text{ij} = 1 \,\, \forall \text{i, j}$). 

\item The action of the second pulse is similar to the action of the first pulse in equation~\eqref{eq: rho^a1}
\begin{equation}
    \boldsymbol{\rho}^\text{a2} \approx \boldsymbol{\rho}^\text{b2} + \frac{i}{\hbar} \left[\boldsymbol{\rho}^\text{b2} + \frac{i}{\hbar} \left[\boldsymbol{\rho}^\text{b2}, \mathbf{V}_2\right], \mathbf{V}_2\right].
\end{equation}

\item Another application of equation~\eqref{eq: free_evolution} gives the density matrix at the temporal position of the  
photon echo $2 \tau_{12}$ 
\begin{equation}
    \rho^\text{PE}_\mathrm{ij} = \rho^\text{a2}_\mathrm{ij} \exp\left\{ i \frac{\tau_{12}}{\hbar} \left[\mathbf{1}, \mathbf{H}_0 \right]_\mathrm{ij} \right\}. 
\end{equation}

\item Finally, we have to extract the components of the density matrix that fulfil the phase matching condition $\overrightharp{k}_\text{PE} = 2\overrightharp{k}_2 - \overrightharp{k}_1$.

\item \label{step_dephasing} Optionally, we can expand the model to account for decay of the 
non-diagonal components of the density matrix, i.e. the optical coherences. 
For that purpose, we introduce phenomenological decoherence times $T_{2, \text{ij}}$ via 
\begin{equation}
    \frac{\text{d}\boldsymbol{\rho}}{\text{d}t} =  \frac{i}{\hbar}\left[\boldsymbol{\rho}, \, \boldsymbol{H}\right] + \boldsymbol{\Gamma}\boldsymbol{\rho} 
    ,\quad \left(\boldsymbol{\Gamma}\boldsymbol{\rho}  \right)_\mathrm{ij} = - \frac{\boldsymbol{\rho}_\mathrm{ij}}{T_{2, \mathrm{ij}}}, \quad \mathrm{i} \neq \mathrm{j}.
\end{equation}
Where $\boldsymbol{\Gamma}$ is the operator that describes the decoherence of the system. 
\end{enumerate}

To automate the modeling procedure described in this Section, we used the 
\href{https://www.python.org/}{Python} library \href{https://www.sympy.org/en/index.html}{SymPy}, 
which resembles other computer algebra systems like Mathematica. 
The source code is available on reasonable \href{mailto:stefan.grisard@tu-dortmund.de}{request}.

\subsection{Exciton-biexciton model}
\begin{landscape}
    \begin{figure}
        \centering 
\includegraphics[scale = 1]{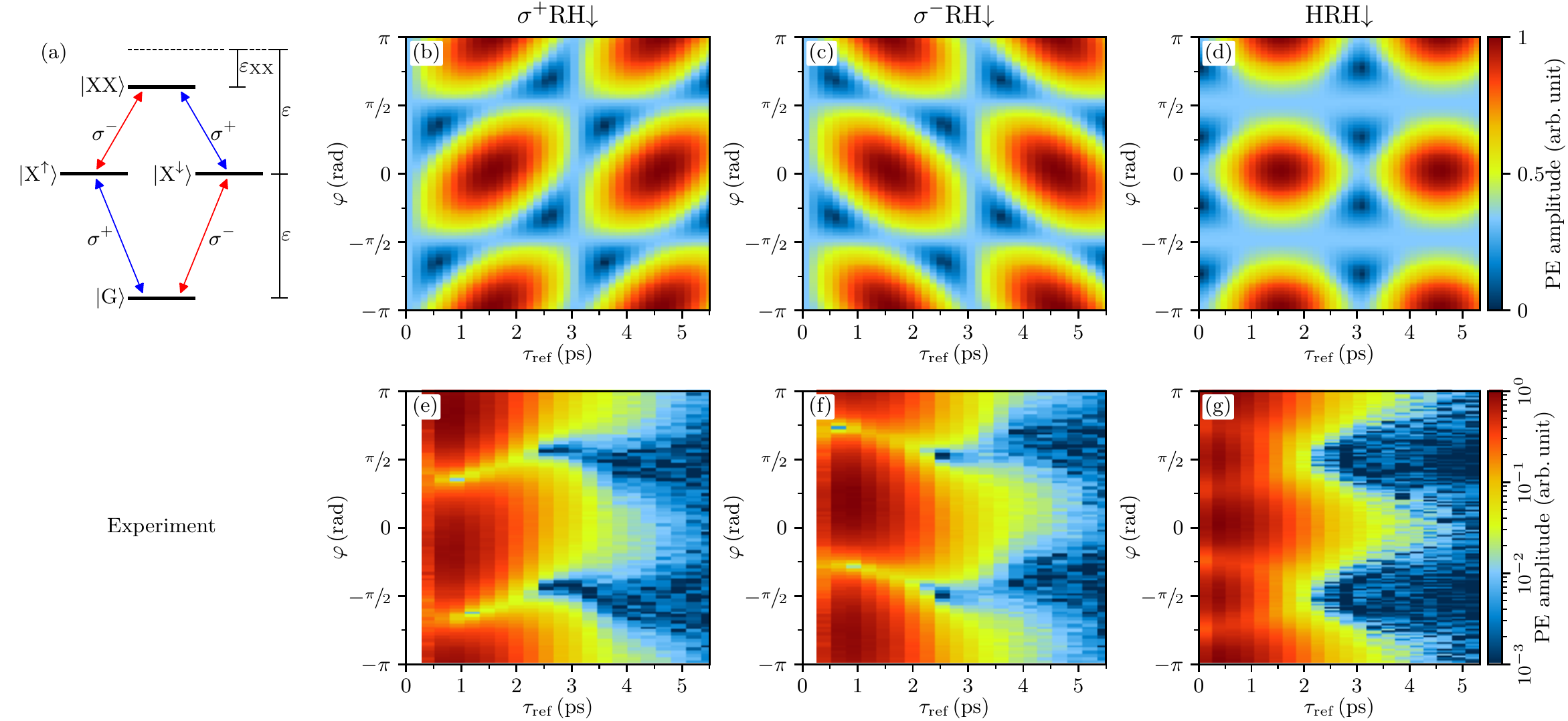}
        \caption{
            Comparison between experiment and the results of the exciton-biexciton model as presented in this Section. 
            The results are arranged in a table.
        The labels on top of each column indicate the polarization configuration.
        In the upper row, we show the results for the exciton-biexciton system with the level-scheme as 
        displayed in (a).
        The lower row shows the experimental colormaps in the three different polarization configurations.}
        \label{fig: biexciton_experiment}
    \end{figure}    
\end{landscape}

As an example for the procedure described above, 
we consider the exciton-biexciton system with a diamond-like level scheme, see Figure~\sref{fig: biexciton_experiment}{a}. 
The Hamilton operator reads as 
\begin{equation}
    \mathbf{H}_0 = 
     \left(\begin{matrix}
        0 & 0 & 0 & 0 \\
        0 & \varepsilon & 0 & 0 \\
        0 & 0 & \varepsilon & 0 \\ 
        0 & 0 & 0 & 2\varepsilon - \varepsilon_\text{XX} \\
     \end{matrix}\right),      
\end{equation}
with the exciton transition energy $\varepsilon$ and the biexciton binding energy $\varepsilon_\text{XX}$.

Taking into account the dipole selection rules as 
indicated by the arrows in Figure~\sref{fig: biexciton_experiment}{a}, the matrix $\mathbf{V}_\mathrm{i}$ corresponding  
to the interaction of the system with the first/second pulse is given by 
\begin{equation}
    \mathbf{V}_\mathrm{i} = d
    \left(\begin{matrix}
        0 & E^*_{+, \mathrm{i}} & E^*_{-, \mathrm{i}} & 0 \\ 
        E_{+, \mathrm{i}} & 0 & 0 & E^*_{-, \mathrm{i}} \\ 
        E_{-, \mathrm{i}} & 0 & 0 & E^*_{+, \mathrm{i}} \\ 
        0 & E_{-, \mathrm{i}} & E_{+, \mathrm{i}} & 0
    \end{matrix}\right),
\end{equation}
with the right- (left-)handed component of the $\mathrm{i}$-th pulse's electric field $E_{+, \mathrm{i}}$ ($E_{-, \mathrm{i}}$). 
For simplicity, we assumed a constant, real dipole matrix element $d$ for all transitions. 
An arbitrary polarization can be expressed in terms of $E_{+, \mathrm{i}}$ and $E_{-, \mathrm{i}}$. For that, 
we introduce the following definitions 
\begin{align}
\begin{aligned}
    E_{+, \mathrm{i}} &= E_{0, \mathrm{i}} \cos\delta_\mathrm{i} \exp\left({+i \varphi_\mathrm{i}}\right)\exp\left({i k_\mathrm{i}}\right) \\
    E_{-, \mathrm{i}} &= E_{0, \mathrm{i}} \sin\delta_\mathrm{i} \exp\left({-i \varphi_\mathrm{i}}\right)\exp\left({i k_\mathrm{i}}\right).
\end{aligned}
\end{align}
In the most general case of elliptically polarized light, $\delta_\mathrm{i}$ measures the ellipticity and 
$\varphi_\mathrm{i}$ the angle between the main principle axis of the ellipse 
and the $x$-axis (horizontal polarization). 
In the following, we consider the polarization configuration $\sigma^+$RH, for which $\delta_1 = 0$, 
$\varphi_1 = 0$ and $\delta_2 = \sfrac{\pi}{4}$. The angle $\varphi_2 \equiv \varphi$ remains a variable 
of the calculation. 
Before the action of first pulse, the system is completely in the ground state, hence 
\begin{equation}
    \boldsymbol{\rho}^\text{b1} = \left(\begin{matrix}
        1 & 0 & 0 & 0 \\
        0 & 0 & 0 & 0 \\
        0 & 0 & 0 & 0 \\ 
        0 & 0 & 0 & 0 \\
     \end{matrix}\right).   
\end{equation}
Following the steps described above, we calculate the density matrix $\boldsymbol{\rho}_{PE}$
\begin{equation}
    \boldsymbol{\rho}^{PE} = - i E_{0, 1} E_{0, 2}^{2} e^{-i (2k_2 - k_1)}
    \left(\begin{matrix}0 &  e^{- i 2 \varphi} &  1 & 0\\
        0 & 0 & 0 & - e^{- i\frac{\varepsilon_\text{XX}}{2\hbar}\tau_\text{ref} }\\
        0 & 0 & 0 & 0\\
        0 & 0 & 0 & 0\end{matrix}
        \right) + \text{adjoint}.
        \label{eq: rho_PE}
\end{equation}
According to equation~\eqref{eq: expectation_value}, the density matrix~\eqref{eq: rho_PE} 
causes right- and left-handed components $P_\pm$ of the photon echo signal. These are
\begin{align}
    \begin{aligned}
    P_+ &\sim \boldsymbol{\rho}^{PE}_{12} + \boldsymbol{\rho}^{PE}_{34} \sim e^{-i 2 \varphi} \\
    P_- &\sim \boldsymbol{\rho}^{PE}_{13} + \boldsymbol{\rho}^{PE}_{24}\sim 1 - e^{-i\frac{\varepsilon_\text{XX}}{2\hbar}\tau_\text{ref}}.
    \label{eq: PEpm}
    \end{aligned}
\end{align}
The final signal in the $\sigma^+$RH configuration can be obtained by 
projection on the horizontal axis
\begin{equation}
    P^{\text{XX}}_{\sigma^+\text{RH}} \sim \left|P_+ + P_- \right|, 
    \label{eq: projection_H}
\end{equation}
which is the special case of the projection on some arbitrary 
detection axis tilted by an angle $\phi$ to the $x$-axis 
\begin{equation}
    P \sim \left|P_+e^{-i \phi}  + P_- e^{+i \phi}  \right|.
\end{equation}
Inserting the expressions~\eqref{eq: PEpm} into~\eqref{eq: projection_H} delivers
\begin{equation}
    P^{\text{XX}}_{\sigma^+\text{RH}} \sim \sqrt{4 \cos^2{\left(\varphi \right)} - 2 \cunderline{red}{\cos{\left(\frac{\tau_\text{ref} \varepsilon_\text{XX}}{2\hbar} \right)}} - 2 \cos{\left(2 \varphi - \cunderline{blue}{\frac{\tau_\text{ref} \varepsilon_\text{XX}}{2\hbar}} \right)} + 1}.
    \label{eq: SpRH_no_damping}
\end{equation}
Here, it becomes clear that 
the signal from a biexciton shows oscillations in the amplitude (second, red underlined term) and 
also a rotation of the 
polar rosettes as a function of time (third, blue). Both are manifestations of quantum beats between 
the exciton and biexciton states. The function \eqref{eq: SpRH_no_damping} is visualized 
in Figure~\sref{fig: biexciton_experiment}{b}.

In the same manner as for the configuration $\sigma^+$RH, we model the signal from the exciton-biexciton system in the configurations 
$\sigma^-$RH and HRH, Figures~\sref{fig: biexciton_experiment}{c} and~\sref{fig: biexciton_experiment}{d}. 
\begin{align}
    P^{\text{XX}}_{\textcolor{red}{\sigma^-}\text{RH}} &\sim \sqrt{4 \cos^2{\left(\varphi \right)} - 2 {\cos{\left(\frac{\tau_\text{ref} \varepsilon_\text{XX}}{2\hbar} \right)}} - 2 \cos{\left(2 \varphi \textcolor{red}{+} {\frac{\tau_\text{ref} \varepsilon_\text{XX}}{2\hbar}} \right)} + 1} \\
    P^{\text{XX}}_\text{HRH} &\sim  \sqrt{8 \smash{\underbrace{\cos^2{\left(\varphi \right)}}_{\text{Two-leaves}}}\sin^2{\left(\frac{\varepsilon_\text{XX}}{4\hbar}\tau_\text{ref} \right)} +  \smash{\underbrace{\cos^2{\left(2 \varphi \right)}}_{\text{Four-leaves}}} } \label{eq: HRH_nodamping}
\end{align}    
The expressions for $\sigma^+$ and $\sigma^-$ excitation 
only differ by the sign of the third term, marked in red. Consequently, the rotational behaviour of the 
signal changes sign upon reversal of helicity of the first pulse 
(compare Figures~\sref{fig: biexciton_experiment}{b} and~\sref{fig: biexciton_experiment}{c}), 
which is a characteristic property of the biexciton diamond-scheme.
In the HRH configuration, the signal is made up of an oscillating part with two-leaves behaviour (i.e. two maxima in the range $\varphi \in [-\pi, \, \pi]$)
and a temporally constant part with four-leaves behaviour (i.e. four maxima in the range $\varphi \in [-\pi, \, \pi]$). 

\clearpage

\section{Analytical expressions for exciton-biexciton system}
In this Section we collect further analytical expressions that were calculated for the exciton-biexciton model. Here 
we also took into account the decoherence time of the exciton and biexciton polarizations. In this way, we obtain 
expressions for the photon echo that serve as fitting functions to our experimental data. This enables us 
to formulate a fitting procedure for estimation of the biexciton binding energy and the decoherence 
times of exciton and biexciton (Section~\ref{sec: fitting_biexciton}).

\subsection{Photon echo amplitude in HRH polarization configuration}
Following step \ref{step_dephasing} on page~\pageref{step_dephasing},
we expand equation~\eqref{eq: HRH_nodamping} to account for decoherence 
of the exciton and biexciton states (decoherence times are denoted as $T_2^\text{X}$ and $T_2^\text{XX}$, respectively)
\begin{equation}
    P^{\text{XX}}_\text{HRH} \sim \sqrt{4\cos^4\left(\varphi\right) e^{-\frac{2\tau_\text{ref}}{T_\text{X}}} +  e^{-\frac{2\tau_\text{ref}}{T_\text{XX}}}  -  4 \cos^2\left(\varphi\right)\cos\left(\frac{\varepsilon_\text{XX}}{2\hbar}\tau_\text{ref}\right) e^{-\tau_\text{ref}\left(\frac{1}{T_\text{X}} + \frac{1}{T_\text{XX}}\right)}}.
\end{equation}    
Two special cases of this equation are the configurations HHH ($\varphi = 0$) and HVH ($\varphi = \sfrac{\pi}{2}$)
\begin{align}
    P^{\text{XX}}_\text{HHH} &\sim \sqrt{4 e^{-\frac{2\tau_\text{ref}}{T_2^\text{X}}} + e^{- \frac{2 \tau_\text{ref}}{T_2^\text{XX}}} - 4\cos{\left(\frac{\varepsilon_\text{XX}}{2\hbar}\tau_\text{ref}\right)}e^{- \frac{\tau_\text{ref}}{T_2^\text{XX}} -  \frac{ \tau_\text{ref}}{T_2^\text{X}}} } \label{eq: HHH} \\
    P^{\text{XX}}_\text{HVH} &\sim e^{- \frac{\tau_\text{ref}}{T_2^\text{XX}}}.\label{eq: HVH}
\end{align}

\subsection[Photon echo amplitude in \texorpdfstring{${\sigma}^\pm$RH}{SRH} polarization configuration]{Photon echo amplitude in \texorpdfstring{$\boldsymbol{\sigma}^\pm$RH}{SRH} polarization configuration}
Following step \ref{step_dephasing} on page~\pageref{step_dephasing},
we expand equation~\eqref{eq: SpRH_no_damping} to account for decoherence 
of the exciton and biexciton states (decoherence times are denoted as $T_2^\text{X}$ and $T_2^\text{XX}$, respectively)
\begin{equation}
    \resizebox{.9\textwidth}{!}{$P^{\text{XX}}_{\sigma^\pm\text{RH}} \sim \sqrt {4\cos^2\left(\varphi\right) e^{-2 \frac{\tau_\text{ref}}{T_2^\text{X}}}
    + e^{-\frac{2\tau_\text{ref}}{T_2^\text{XX}}}
    - 2 \left[\cos\left(\frac{\varepsilon_\text{XX}}{2\hbar}\tau_\text{ref}\right) + \cos\left(2\varphi \mp \frac{\varepsilon_\text{XX}}{2\hbar} \tau_\text{ref}\right)\right]e^{-\tau_\text{ref}\left(\frac{1}{T_2^\text{X}} + \frac{1}{T_2^\text{XX}}\right)}}.$}
    \label{eq: SpRH}
\end{equation}

\subsection[Comparison of \texorpdfstring{${\sigma}^\pm$HR}{SHR} and \texorpdfstring{${\sigma}^\pm$RH}{SRH}]{Comparison of \texorpdfstring{$\boldsymbol{\sigma}^\pm$HR}{SHR} and \texorpdfstring{$\boldsymbol{\sigma}^\pm$RH}{SRH}}

\begin{figure}
    \centering
\includegraphics[scale = 1]{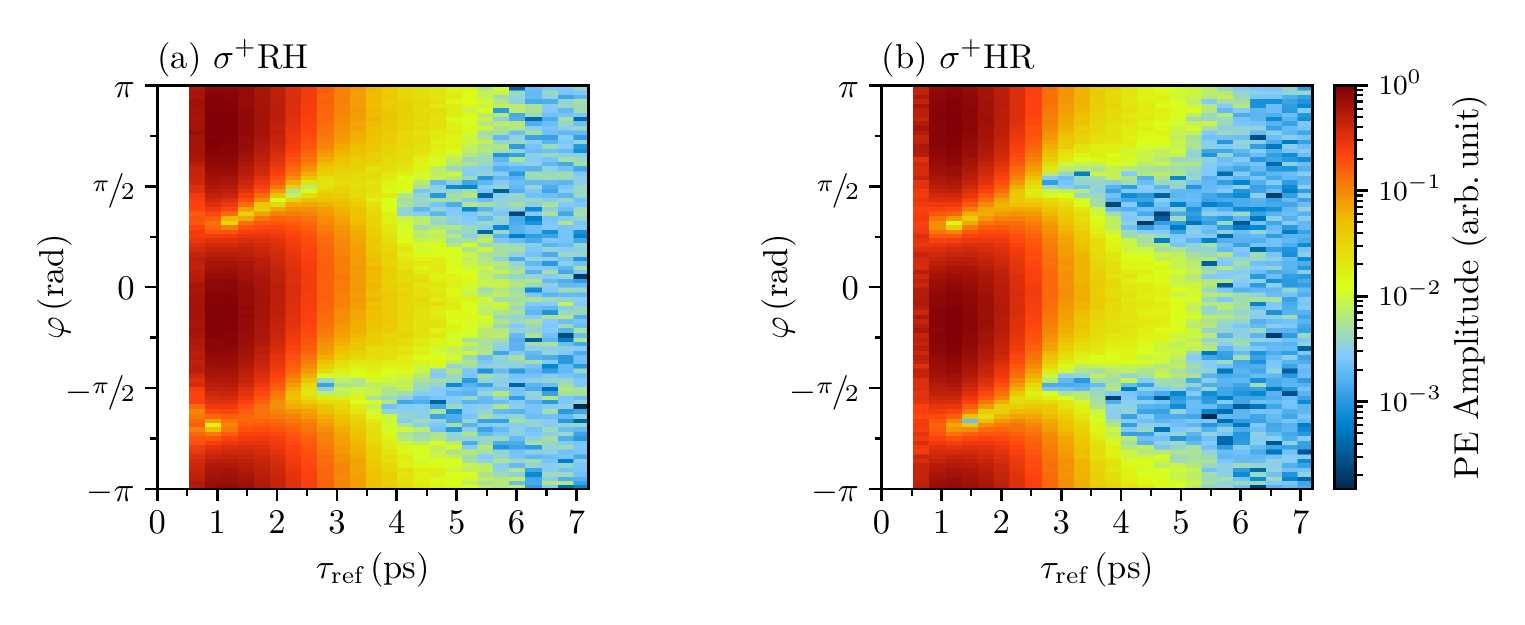}
    \caption{Experimental comparison between the configurations $\sigma^{\!+}$RH and $\sigma^{\!+}$HR. The data sets 
    are not distinguishable, which is in agreement with the prediction of equation~\eqref{eq: SPRR}.}
    \label{fig: SpRH_SPHR}
\end{figure}

In this Section, we discuss the role of the polarization 
of the second pulse on the dynamics on the Poincaré sphere. For that purpose, it is illuminating 
to calculate the photon echo signal as a function of the polarization angle of the second pulse $\varphi$
and the detection angle $\phi$
\begin{equation}
    P^{\text{XX}}_{\sigma^{\!+}\varphi\phi} \sim \sqrt{4 \cos^2{\left(\varphi  + \phi\right)} - 2 {\cos{\left(\frac{\tau_\text{ref} \varepsilon_\text{XX}}{2\hbar} \right)}} - 2 \cos{\left(2 \varphi  + 2\phi - {\frac{\tau_\text{ref} \varepsilon_\text{XX}}{2\hbar}} \right)} + 1}.
    \label{eq: SPRR}
\end{equation}
Here it becomes clear, that the polarization of the second pulse and the detection angle are completely 
exchangeable. For example, the configurations $\sigma^{\!+}$RH and $\sigma^{\!+}$HR result in the same signal, 
as we observe in experiment (Figure~\ref{fig: SpRH_SPHR}). 

\begin{figure}
    \centering
\includegraphics[scale = 1]{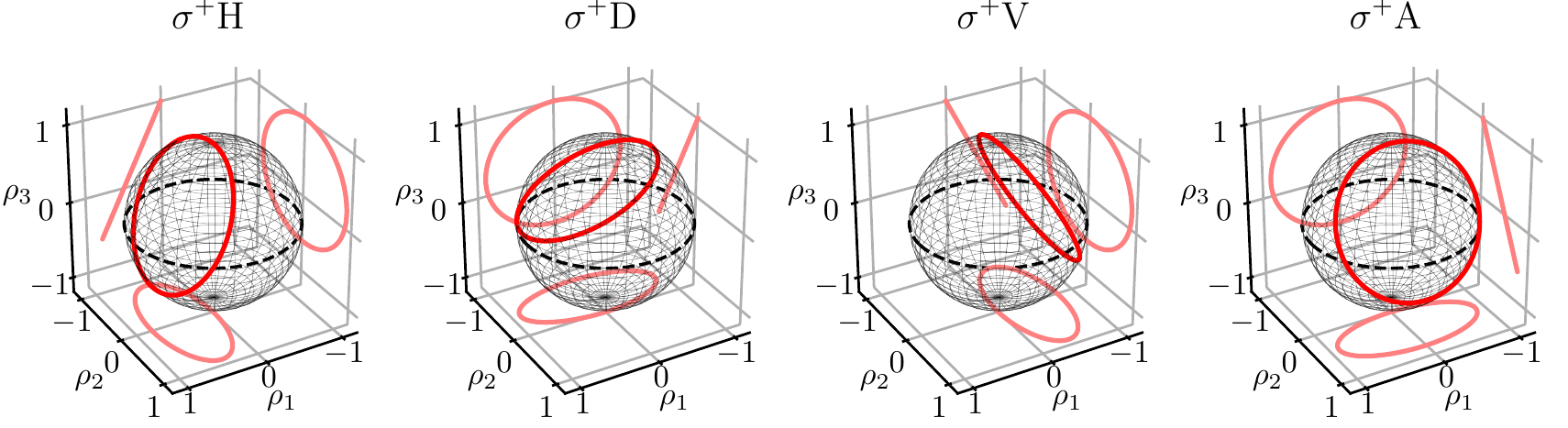}
    \caption{Visualization of the effect of the second pulse's polarization on the 
    PE's polarization state on the Poincaré sphere.}
    \label{fig: poicare_spheres_SpRR}
\end{figure}

To understand the effect of the 
second pulse's polarization 
on the dynamics on the Poincaré sphere, 
we show in Figure~\ref{fig: poicare_spheres_SpRR} the trajectories for different 
polarizations of the second pulse. Here, we can see that rotation of the second 
pulse leads to a rotation of the trajectory about the $\rho_3$ axis.
Indeed, we can write the analytical expression for the Stokes vector as a function of 
the second pulse's linear polarization $\varphi$ as 
\begin{equation}
    \left(\begin{matrix}
        \rho_1^{\sigma^{\!+}}(\varphi)\\
        \rho_2^{\sigma^{\!+}}(\varphi)\\
        \rho_3^{\sigma^{\!+}}(\varphi)
    \end{matrix}\right)= 
    \underbrace{
    \left(\begin{matrix}
        \cos(2\varphi) & \sin(2\varphi) & 0 \\
        -\sin(2\varphi) & \cos(2\varphi) & 0 \\
        0 & 0 & 1 \\
    \end{matrix}\right) }_{\text{Rotation by $2\varphi$ about $\rho_3$ axis}}
    \left(\begin{matrix}
        \rho_1^{\sigma^{\!+}\text{H}}\\
        \rho_2^{\sigma^{\!+}\text{H}}\\
        \rho_3^{\sigma^{\!+}\text{H}}
    \end{matrix}\right),
    \label{eq: rotation}
\end{equation}
where 
\begin{align}
    \rho_1^{\sigma^{\!+}\text{H}} &= \frac{2\cos\left(\frac{\varepsilon_\text{XX}\tau_\text{ref}}{2\hbar}\right) - 2}{2\cos\left(\frac{\varepsilon_\text{XX}\tau_\text{ref}}{2\hbar}\right) - 3},\\
    \rho_2^{\sigma^{\!+}\text{H}} &= \frac{-2\sin\left(\frac{\varepsilon_\text{XX}\tau_\text{ref}}{2\hbar}\right)}{2\cos\left(\frac{\varepsilon_\text{XX}\tau_\text{ref}}{2\hbar}\right) - 3}, \\
    \text{and}\,\, \rho_3^{\sigma^{\!+}\text{H}} &= \frac{1 - 2\cos\left(\frac{\varepsilon_\text{XX}\tau_\text{ref}}{2\hbar}\right)}{2\cos\left(\frac{\varepsilon_\text{XX}\tau_\text{ref}}{2\hbar}\right) - 3} 
\end{align}
are the Stokes parameters in the configuration $\sigma^{\!+}$H ($\varphi = 0$).

\clearpage
\section{Description of fitting procedure}
In this Section we describe in detail how we obtained values for the biexciton binding energy, 
decoherence times of exciton and biexciton, and the (in)homogeneous linewidth of the studied ensemble
from fits to our experimental data. 

\subsection{Fitting procedure for evaluation of biexciton binding energy and decoherence times of exciton and biexciton}
\label{sec: fitting_biexciton}

\begin{figure}
    \centering 
\includegraphics[scale = 1]{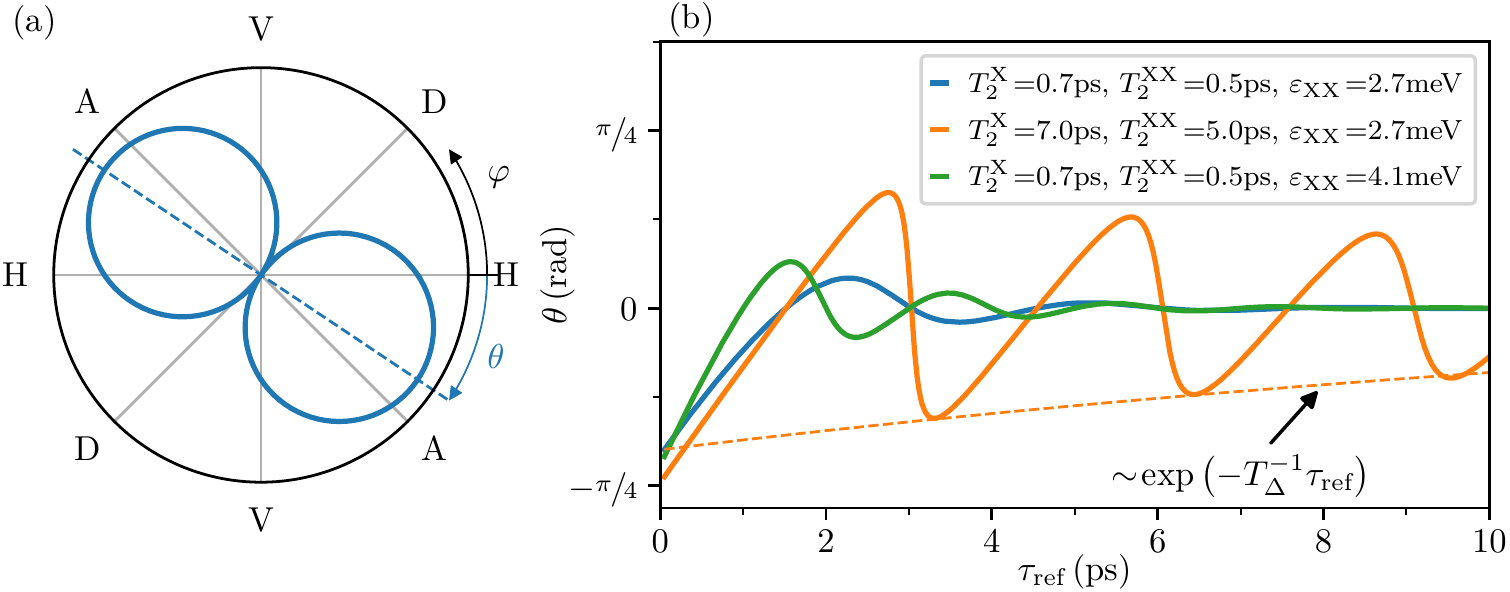}
    \caption{a) Definition of the angle $\theta$ as the angle between the H-axis and 
    the main axis of a two-leave polar rosette. b) Visualization of equation~\eqref{eq: vartheta_offset}
    for different sets of values $\{T_2^\text{X}$, $T_2^\text{XX}$, $\varepsilon_\text{XX}\}$.}
    \label{fig: theta}
    \end{figure}  

As mentioned earlier, quantum beats between exciton and biexciton can be observed 
in the polarization configuration $\sigma^+$RH. In contrast to the polarization configuration 
HRH, the quantum beats not only modulate the amplitude of the photon echo signal, but also the 
orientation of polar rosettes (a direct manifestation of polarizations beats). 
As will be become clear below, the polarization beating 
offers us the possibility to extract the biexciton binding energy with high significance although the decoherence times of 
exciton and biexciton are too short to observe amplitude beats. 
  
To quantify the rotation of the rosettes, 
we introduced in the main text the parameter $\theta$ that measures the angle between the principle 
axis of the rosette 
and the H-axis (see definition in Figure~\sref{fig: theta}{a}). 
For a fixed value of $\tau_\text{ref} > 0$, $\theta$ is equal to the angle $\varphi \in [-\sfrac{\pi}{2}, \sfrac{\pi}{2}]$ 
that maximizes the signal $P^\text{XX}_{\sigma^\pm\text{RH}}$. 
To find an analytical expression for $\theta$ as function of $\tau_\text{ref}$, we 
set the derivative of equation~\eqref{eq: SpRH} equal to zero: 
\begin{align}
\frac{\mathrm{d}}{\mathrm{d}\varphi} P^{\text{XX}}_{\sigma^\pm\text{RH}} & = 0 \label{eq: 0} \\ 
\Leftrightarrow 0 &= \sin\left( 2\varphi \mp \frac{\varepsilon_\text{XX}}{2\hbar} \tau_\text{ref}\right) e^{-\tau_\text{ref}\left(\frac{1}{T_\mathrm{X}} + \frac{1}{T_\mathrm{XX}}\right)} - \sin\left(2 \varphi\right)e^{-2 \frac{\tau_\text{ref}}{T_\mathrm{X}}} \\
\Leftrightarrow \varphi &= \pm\frac{1}{2}\arctan\left[\frac{\sin\left(\frac{\varepsilon_\text{XX}\tau_\text{ref}}{2\hbar}\right)}{\cos\left(\frac{\varepsilon_\text{XX}\tau_\text{ref}}{2\hbar}\right) - e^{\tau_\text{ref}\left(\frac{1}{T_\mathrm{XX}} - \frac{1}{T_\mathrm{X}}\right) }}  \right] \equiv \theta_{\sigma^\pm}(\tau_\text{ref}),  \label{eq: transcendental}
\end{align}	
where the index $\sigma^\pm$ in the definition of the function $\theta(\tau_\text{ref})$ 
distinguishes between the polarization configurations $\sigma^+$RH and $\sigma^-$RH.
Figure~\sref{fig: theta}{b} visualizes $\theta_{\sigma^+}(\tau_\text{ref})$
for three different sets of values for $T_2^\text{X}$, $T_2^\text{XX}$, and $\varepsilon_\text{XX}$.

A remarkable property of equation~\eqref{eq: transcendental} is that the 
temporal decay constant in the nominator of the argument of the 
arctangent is given by the \emph{difference} $T_\Delta^{-1}$ between 
the decay rates of biexciton and exciton
\begin{equation} 
    \frac{1}{T_\Delta} \equiv \frac{1}{T^\text{XX}_2} - \frac{1}{T^\text{X}_2} > 0.
    \label{eq: definition_gamma}
\end{equation}
Therefore, the envelope of the function~\eqref{eq: transcendental}
decays proportionally to $\exp(-\sfrac{\tau_\text{ref}}{T_\Delta})$, which is 
exemplarily highlighted in Figure~\sref{fig: theta}{b} by the dashed line for the orange curve. 
This property is the main reason why the study of polarization beats, rather than amplitude 
quantum beats enables us to obtain the biexciton binding energy with high significance. 
In contrast, the amplitude quantum beats, for example measurable in the configuration HHH (see equation~\eqref{eq: HHH}), 
decay with a rate given by the \emph{sum} of the decay rates of exciton and biexciton. 
Since the sample gives rise to short decoherence times, it is not possible to observe 
amplitude quantum beats. 

\begin{figure}
    \centering 
\includegraphics[scale = 1]{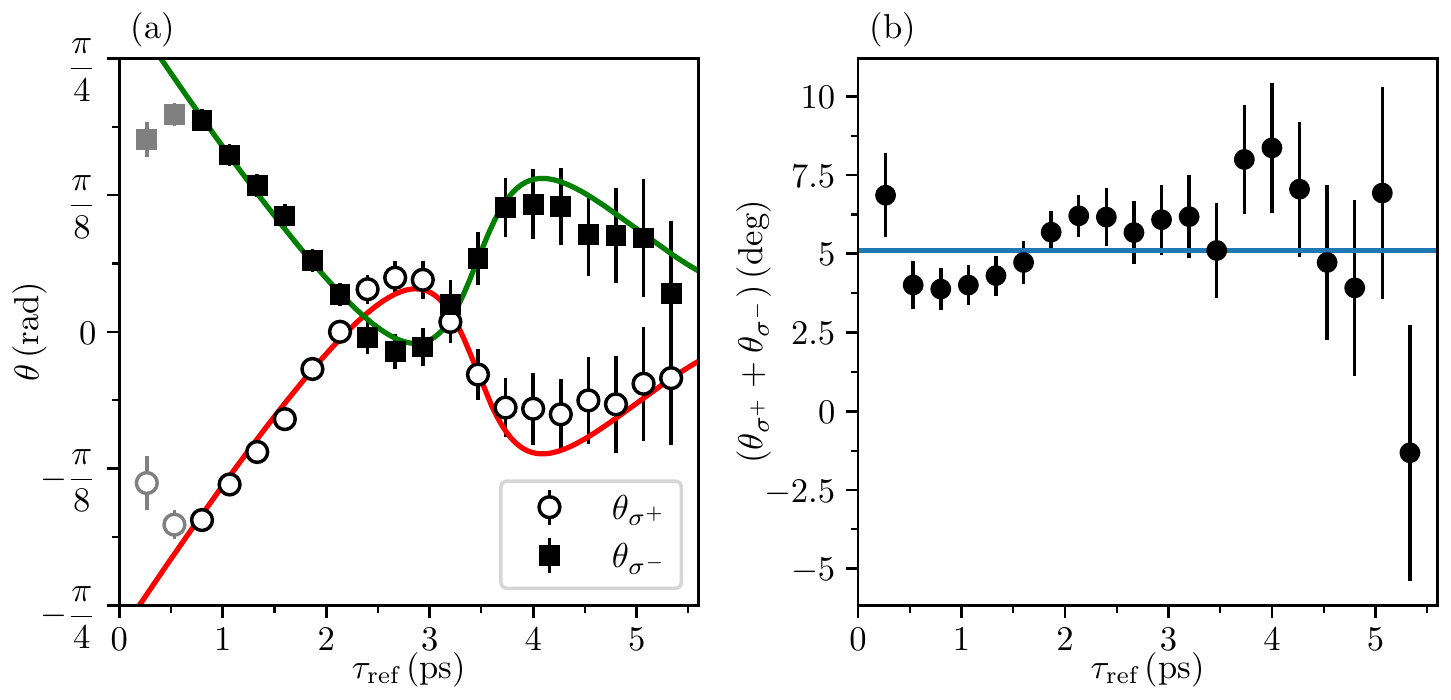}
    \caption{Experimental data for $\theta (\tau_\text{ref})$ in the configurations $\sigma^+$RH and $\sigma^-$RH and corresponding 
    fits to equation~\eqref{eq: vartheta_offset}.}
    \label{fig: theta_shift}
\end{figure}
\begin{figure}
\centering 
\includegraphics[scale = 1]{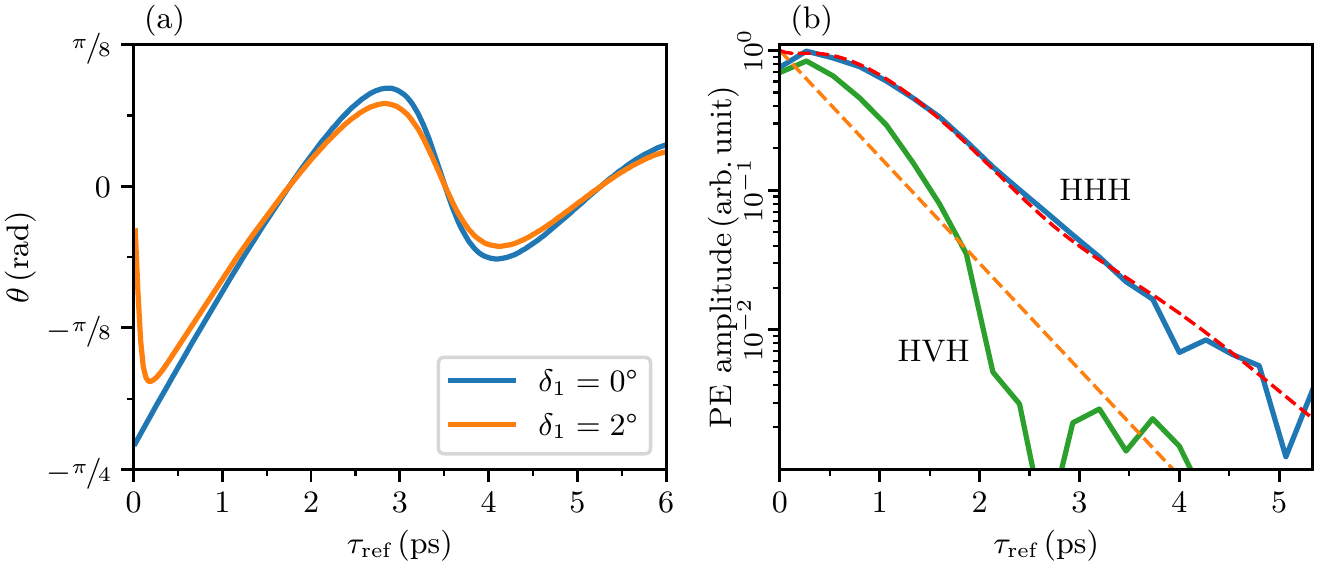}
\caption{(a) Comparison of modeled functions $\theta(\tau_\text{ref})$ for two different values 
of ellipticity $\delta_1$ of the first pulse. (b) Experimental and modeled decay of the PE amplitude in the 
polarization configurations HHH and HVH.}
\label{fig: elliptical}
\end{figure}

Equation~\ref{eq: transcendental} predicts that the function $\theta(\tau_\text{ref})$ is mirrored on 
the $\tau_\text{ref}$-axis upon change of helicity of the first pulse's polarization, 
i.e. $\theta_{\sigma^-}(\tau_\text{ref}) = -\theta_{\sigma^+}(\tau_\text{ref})$. Consequently, the sum 
of both curves vanishes. However, in experiment we observe an  
offset of roughly $\SI{5}{\degree}$ of the sum of both curves, which is shown in Figure~\sref{fig: theta_shift}{b}. The blue line 
corresponds to a fit to a constant function. 
To account for this discrepancy between model and experiment, we phenomenologically expand the model~\ref{eq: transcendental} 
by an offset $\theta_0^{\pm}$
\begin{equation}
    \theta_{\sigma^\pm}(\tau_\text{ref}) = \pm\frac{1}{2}\arctan\left[\frac{\sin\left(\frac{\varepsilon_\text{XX}\tau_\text{ref}}{2\hbar}\right)}{\cos\left(\frac{\varepsilon_\text{XX}\tau_\text{ref}}{2\hbar}\right) - e^{\tau_\text{ref}\left(\frac{1}{T_\mathrm{XX}} - \frac{1}{T_\mathrm{X}}\right) }}  \right] + \theta_0^\pm.
    \label{eq: vartheta_offset}
\end{equation}
Furthermore, the comparison of the experimental data 
for $\theta_{\sigma^\pm} = \theta(\tau_\text{ref})$ in Figure~\sref{fig: theta_shift}{a}
with the modeled curves in Figure~\sref{fig: theta}{b} reveals that the model does not adequately describe 
the experiment in the range $\tau_\text{ref} < \SI{0.8}{\pico\second}$. In particular, the first two data points 
(colorized in gray) deviate 
from the linear trend of the modeled functions. 
This effect can be caused by two contributions.
First, our model neglects the temporal overlap of the optical pulses that takes place in the range 
$\tau_\text{ref} < \SI{0.8}{\pico\second}$ (see figure 1(c) in the main text). 
Second, our model assumes perfect circular polarization of the first pulse. However, in our experimental scheme,
the circular polarization that is created by a quarter wave plate could be altered by the subsequent reflection 
on two silver mirrors. 
In Figure~\ref{fig: elliptical} we numerically calculated 
$\theta(\tau_\text{ref})$ for an elliptical polarized first pulse. 
The behaviour for small values of $\tau_\text{ref}$ resembles the experimental observations. 
However, for elliptically polarized light there exists no closed solution for 
$\theta(\tau_\text{ref})$, which makes fitting computationally 
expensive and impractical using standard fitting tools. 
Hence, we adhere to the simlified function~\eqref{eq: vartheta_offset} to fit our data and, for simplicity, do not 
include the first two data points in Figure~\sref{fig: theta_shift}{b}. 

To obtain a value for $\varepsilon_\text{XX}$, we combine 
the data for $\theta_{\sigma^+}(\tau_\text{ref})$ and $\theta_{\sigma^-}(\tau_\text{ref})$ 
by fitting $(\theta_{\sigma^+} - \theta_{\sigma^-}) / 2$ for 
$\tau_\text{ref} > \SI{0.8}{\pico\second}$ to equation~\eqref{eq: vartheta_offset}. 
Additionally, the fit gives us a value for the difference of 
decay rates $T_\Delta^{-1}$ and the difference between the offsets $\theta_0^+ - \theta_0^-$. 
All fitting parameters are summarized in table~\ref{tab: fitting_parameters_theta}.  
\begin{table}
	\centering
	\caption{Summary of the parameters obtained from fit of the model~\eqref{eq: vartheta_offset} to experimental data for
    $\theta(\tau_\text{ref})$ in the combined data of the polarization configurations $\sigma^+$RH and 
    $\sigma^-$RH. The data and corresponding fits are shown in Figure~\ref{fig: theta_shift}.}
	\label{tab: fitting_parameters_theta}
	\begin{tabular}{
		S[table-format=1.1] @{${}\pm{}$} S[table-format=1.1]
		S[table-format=2.2] @{${}\pm{}$} S[table-format=1.2]
		S[table-format=2] @{${}\pm{}$} S[table-format = 1]
        S[table-format=2] @{${}\pm{}$} S[table-format = 1]
		}
	\toprule
		\multicolumn{2}{c}{$\varepsilon_\text{XX}$\,(\si{\milli\eV})} &
        \multicolumn{2}{c}{$T_\Delta^{-1}$\;(\si{\tera\hertz})} & 
        \multicolumn{2}{c}{$\theta_0^+$\;(deg)} &
        \multicolumn{2}{c}{$\theta_0^-$\;(deg)} \\
	\midrule
        2.4 & \!\! 0.2 & 0.23 & 0.04 & -9 & \!\! 1 & 14 & \!\! 1 \\
	\bottomrule
	\end{tabular}
\end{table}

Next, we use the experimental curve of the photon echo decay in the configuration HHH to extract the exciton decoherence 
time $T_2$ (Figure~\sref{fig: elliptical}{b}). 
Here, we want to take into account the obtained values for 
$\varepsilon_\text{XX}$ and $T_\Delta^{-1}$ and leave only $T_2^\text{X}$ as free fitting parameter.
Therefore, we write the function~\eqref{eq: HHH} in terms of $T_\Delta^{-1}$, $\varepsilon_\text{XX}$ and $T_2^\text{X}$:
\begin{align}
    P^{\text{XX}}_\text{HHH} &\sim \sqrt{4 e^{-\frac{2\tau_\text{ref}}{T_2^\text{X}}} + e^{- \frac{2 \tau_\text{ref}}{T_2^\text{XX}}} - 4\cos{\left(\frac{\varepsilon_\text{XX}}{2\hbar}\tau_\text{ref}\right)}e^{- \frac{\tau_\text{ref}}{T_2^\text{XX}} -  \frac{ \tau_\text{ref}}{T_2^\text{X}}} } \\
    &\overset{\eqref{eq: definition_gamma}}{=} e^{-\frac{\tau_\text{ref}}{T_2^\text{X}}} \sqrt{4 + e^{- 2 \frac{\tau_\text{ref}}{T_\Delta}} - 4\cos{\left(\frac{\varepsilon_\text{XX}}{2\hbar}\tau_\text{ref}\right)}e^{- \frac{\tau_\text{ref}}{T_\Delta}} }
\end{align}
The fit gives
\begin{equation}
 T_2^\text{X} = \SI{0.79(3)}{\pico\second} 
 \label{eq: T2X}
\end{equation}
and is visualized in Figure~\sref{fig: elliptical}{b}.
Using the definition of the $T_\Delta^{-1}$, we can calculate $T_2^\text{XX}$
\begin{equation}
    T_2^\text{XX} = \left(\frac{1}{T_\Delta} + \frac{1}{T_2^\text{X}}\right)^{-1}  = \SI{0.67(3)}{\pico\second}.
    \label{eq: T2XX} 
\end{equation}
According to our model, the photon echo signal decays in the HVH configuration proportional to 
$\exp\left(-\sfrac{\tau_\text{ref}}{T_2^\text{XX}}\right)$ (see equation~\eqref{eq: HVH}). The comparison 
between the experimental decay and the model using the obtained value for $T_2^\text{XX}$ is shown in 
Figure~\sref{fig: elliptical}{b}. Here we can see that the assumption of an exponential decay of the biexciton 
coherence is not supported by the experiment. Rather, the signal decays in a Gaussian manner, i.e. proportional 
to $\exp\left(-\sfrac{\tau^2_\text{ref}}{\sigma^2_\text{XX}}\right)$ with a temporal width $\sigma_\text{XX}$.
This effect could be related to an inhomogeneity of the decoherence time $T_2^\text{XX}$.

\subsection{Extraction of homogeneous and inhomogeneous linewidths}
The \emph{homogeneous} spectral broadening of exciton and biexciton 
$\Gamma_\text{hom}^\text{X}$ and $\Gamma_\text{hom}^\text{XX}$
are determined by the decoherence times $T_2^\text{X}$ and $T_2^\text{XX}$ 
that we extracted in the \hyperref[sec: fitting_biexciton]{previous Section}.
Therefore, we assume that the spectral lineshape corresponding to the temporal signal 
$\exp(\text{-}\sfrac{\tau_\text{ref}}{T_2})$ is given by the real part of its Fourier transform
$\frac{\sfrac{1}{T_2}}{\sfrac{1}{T_2^2} + \omega^2}$, which is a Lorentzian 
function with FWHM $\sfrac{2}{T_2}$. 
Expressed in energetic widths, this relationship leads us to
\begin{equation}
    \Gamma_\text{hom}^\text{X} = \frac{2 \hbar}{T_2^\text{X}} \overset{\eqref{eq: T2X}}{=} \SI{1.67(6)}{\milli\eV}, \quad
    \Gamma_\text{hom}^\text{XX} = \frac{2 \hbar}{T_2^\text{XX}} \overset{\eqref{eq: T2XX}}{=} \SI{1.97(8)}{\milli\eV}.
\end{equation}

\begin{figure}
    \centering
\includegraphics[scale = 1]{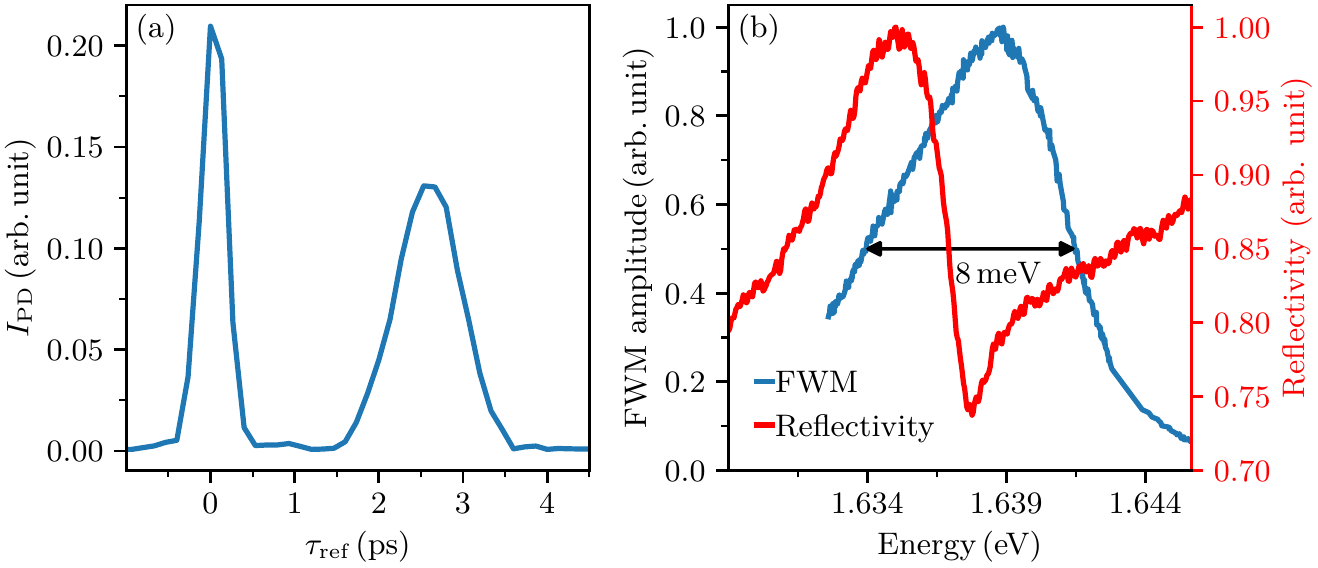}
    \caption{(a) Dependence of the heterodyne signal on $\tau_\text{ref}$ for a fixed 
    value of $\tau_{12} = \SI{1.5}{\pico\second}$. (b) Four-Wave-Mixing spectrum, measured using spectrally narrow 
    laser pulses.}
    \label{fig: spectrum}
\end{figure}    
Next, we want to quantify the \emph{inhomogeneous} broadening that gives rise to the PE effect. 
Therefore, we analyze the time-resolved measurements of the PE signal in Figure~\sref{fig: spectrum}{a}. 
For this measurement, the temporal gap between first and 
second pulse is $\tau_{12} = \SI{1.5}{\pico\second}$. The signal at $\tau_\text{ref} = \SI{0}{\pico\second}$ 
arises from the cross correlation 
of scattered light from the first pulse and the reference. 
This signal serves as a reference to measure the temporal width of the laser pulses. 
In particular, we extract the temporal width of the laser pulses by a fit of the data in 
the range $\tau_\text{ref} < \SI{1}{\pico\second}$
to a Gaussian function of the form
\begin{equation}
    g(\tau_\text{ref}) = A \exp\left\{ -\frac{\left(\tau_\text{ref} - \tau_0 \right)^2}{2\sigma^2}   \right\} = A \exp\left\{ -\frac{4\text{ln}(2)\left(\tau_\text{ref} - \tau_0 \right)^2}{f^2} \right\}, 
\end{equation}     
Here, the amplitude $A$, and the temporal shift $\tau_0$ are fitting parameters. 
In the last step, we substituted the standard deviation $\sigma$ by the full width at half maximum (FWHM) 
$f = 2\sqrt{2\text{ln}(2)}\sigma \approx \num{2.355}\cdot\sigma$, which 
is more commonly used for the discussion of spectral or temporal widths. 
The fit delivers the following width of the cross correlation 
\begin{equation}
    f_\text{CC} = \SI{0.35(12)}{\pico\second}. 
\end{equation} 
Since the cross correlation of two Gaussians with variances $\sigma_1^2$ and $\sigma_2^2$ is itself also 
a Gaussian with variance $\sigma_1^2 + \sigma_2^2$, the temporal width $f_L$ of the laser pulses (related to its elctric field) is
\begin{equation}
    f_L = \frac{f_\text{CC}}{\sqrt{2}} = \SI{0.25(8)}{\pico\second},
    \label{eq: laserduration}
\end{equation}
which corresponds to an intensity width of roughly $\SI{0.18}{\pico\second}$.
To extract the inhomogeneous linewidth, we analyze the temporal width of the PE shown above.
A fit of the data for $\tau_\text{ref} > \SI{1}{\pico\second}$ gives a temporal width of $\SI{0.92(12)}{\pico\second}$. 
Taking into account that the measured signal is the 
cross correlation of the pure PE signal and the reference pulse, we arrive at a temporal width of 
\begin{equation}
    T_2^* = \SI{0.88(12)}{\pico\second}, 
\end{equation}
which corresponds to the inhomogeneous linewidth 
\begin{equation}
    \Gamma_\text{inh} = \frac{8 \ln (2)\hbar}{T_2^*} = \SI{4.1(6)}{\milli\eV}. 
\end{equation}
As a final step, we want to estimate the overall spectral width associated with the exciton ensemble. 
This enables us to compare the results from time-domain 
with the spectral measurements that we present in Figure~\sref{fig: spectrum}{b}. 
The overall spectrum is given by the convolution between the homogeneous line (Lorentzian) 
and the inhomogeneous line (Gaussian), which results in a Voigt lineshape. Given the width of the Lorentzian 
and Gaussian line ($\Gamma_\text{hom}^\text{X}$ and $\Gamma_\text{inh}$), the FWHM of the resulting Voigt line $\Gamma_\text{V}$ 
is approximately determined by\cite{oliveroempirical1977}
\begin{equation}
    \Gamma_\text{V} = \num{0.5346}\cdot\Gamma_\text{hom}^\text{X} + \sqrt{\num{0.2166}\cdot\left(\Gamma_\text{hom}^\text{X}\right)^2 + \Gamma_\text{inh}^{2}} = \SI{5.1(8)}{\milli\eV}.
\end{equation}
This value is by $\SI{2.9(8)}{\milli\eV}$ smaller than the FWHM $\SI{8}{\milli\eV}$ that we observe in the 
spectral~measurement, Figure~\sref{fig: spectrum}{b}. 
This deviation is reasonable taking into account two possible contributions. First, we supposed that the transformation 
between energy- and time-domain is fully determined by the Fourier transform. This assumption may be 
oversimplified for non-ideal, i.e. non-bandwidth-limited, laser pulses. 
Second, we assumed that the laser spectrum for the time-domain measurements is significantly broader than 
the inhomogeneous broadening. In this ideal case, the whole ensemble is excited equally. However, in our 
experiment the width of the amplitude spectrum of the laser $\Gamma_L$ is limited by 
\begin{equation}
    \Gamma_L =  \frac{8 \ln (2)\hbar}{f_L} \overset{\eqref{eq: laserduration}}{\approx} \SI{15}{\milli\eV} \quad \text{(intensity width $\approx\!\SI{11}{\milli\eV}$)},
\end{equation}
and is therefore not essentially broader than the excited ensemble, compare Figure~\sref{fig: spectrum}{b}.

As additional information, we measured the reflectivity spectrum of 
the \ce{MAPbI3} sample, shown together with 
the FWM spectrum in figure~\sref{fig: spectrum}{b}. 
The reflectivity spectrum may contain contributions from the real 
and imaginary part of the linear optical response, which explains the dispersive 
lineshape. 
Nevertheless, we can see that, both, spectral position 
of the resonance and the spectral width of FWM and reflectivity spectrum coincide. 
Note that 
the analysis of reflectivity spectra of bulk materials is a nontrivial task and 
polariton effects as well as surface effects should be taken into account.
We emphasize that our transient four-wave mixing technique provides richer information 
as linear spectroscopic methods. 
In particular, it is possible to separate the contribution from homogeneous and 
inhomogeneous broadening of optical 
transitions as is discussed above.

\clearpage
\newpage
\section{Additional modeling results}\label{sec: modeling_excitonic_complexes}
In this paper, the comparison between the outcome of our theoretical model 
(as presented in Section~\ref{sec: modeling}) 
with experimental observations for various polarization configurations allowed us to unambiguously 
identify the presence of biexcitons in the studied \ce{MAPbI3} sample.   
We have shown that the measurement of polar rosettes as a function of the delay time between first and 
second pulse places stringent constraints on an excitonic model that gives rise to quantum beats. 

As an outlook, we use in the following our model to predict the behaviour of other exciton systems 
in similar experiments. First, we discuss the polarization 
interference between excitons and trions in 
Section~\ref{sec: exciton_trion}. Here we point out how time-resolved FWM can distinguish between 
quantum beats and polarization interference. 
Second, we show the results of the application of our model to a variety of 
excitonic systems that exhibit any kind of energy splitting giving rise to quantum beats. 
Here, we present our measurement protocol as an illuminating tool to characterize 
systems that are not well understood in terms of their energy level structure and polarization 
selection rules. 

\subsection{Exciton-trion polarization interference}\label{sec: exciton_trion}
\begin{landscape}
    \begin{figure}
        \centering 
\includegraphics[scale = 1]{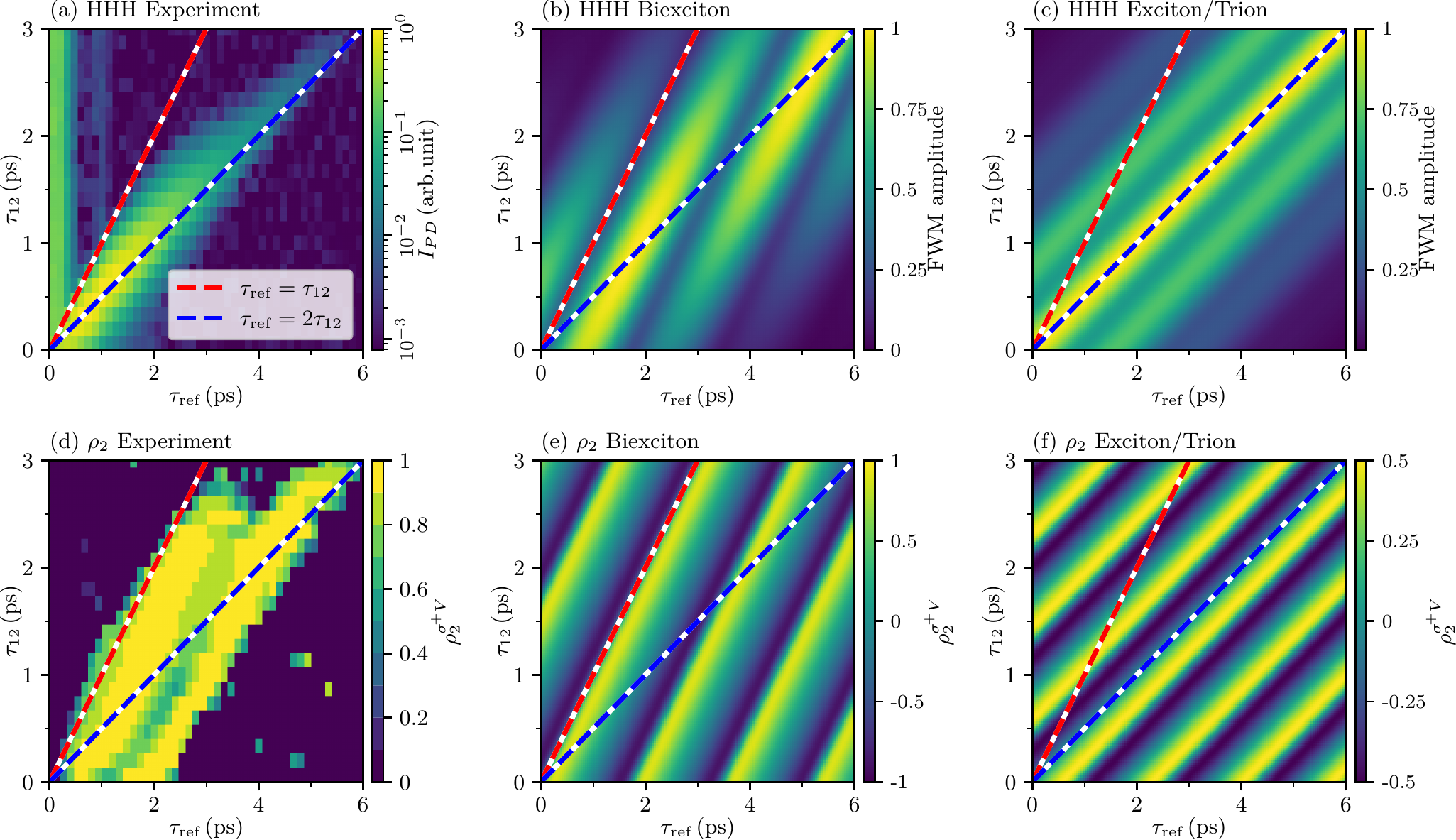}
        \caption{(a) Heterodyne signal as a function of $\tau_{12}$ and $\tau_\text{ref}$, measured in the polarization configuration HHH.
        (b)/(c) Modeled data within the biexciton/exciton+trion model corresponding to (a). For modeling we neglected the finite decoherence times 
        of the involved coherences and, for better visibility of the oscillating signal, assumed a narrow inhomogeneous broadening. 
        (d) Polarization contrast $\rho_2^{\sigma^{\!+}\text{V}}$ as a function of $\tau_{12}$ and $\tau_\text{ref}$. (e)/(f) corresponding modeled 
        data within the two models.}
        \label{fig: QB_PI_discussion}
    \end{figure}    
\end{landscape}

In Figures 2(c) and 2(d) of the main text, we showed that the studied sample gives rise to different types 
of polar rosettes depending on the delay between first and second pulse. One resembles the rosette of 
an exciton (see Figure 1(e)), the other that of a trion (see Figure 1(f)). 
A natural solution for this observation is the independent excitation of noninteracting excitons and trions, giving rise to 
polarization interference (PI). 
However, the transient four-wave-mixing technique allows to distinguish unambiguously between quantum beats (QB) and 
polarization interference by measurement of the FWM signal as a function 
of $\tau_{12}$ and $\tau_\text{ref}$, as was pointed out first by Koch et al.\cite{Koch-PRL1992}.
To visualize the different behavior 
of QB and PI within this experimental protocol, 
we modelled in Figures~\sref{fig: QB_PI_discussion}{b} and \hyperref[fig: QB_PI_discussion]{(c)} the PE signal 
in the HHH polarization configuration from biexciton 
and exciton/trion, respectively. Figure~\sref{fig: QB_PI_discussion}{a} shows the corresponding 
experimental data for comparison. For the models, we assumed a biexciton 
binding energy of 
$\varepsilon_\text{XX} = \SI{2.4}{\milli\eV}$ and 
a splitting $\Delta \varepsilon$ between exciton and trion transition of the same magnitude.  
To see several oscillation cycles in the theoretical colormaps 
within the temporal width of the photon echo, we assumed a smaller inhomogeneous broadening 
of roughly $\SI{2}{\milli\eV}$ than in experiment ($\SI{4.1}{\milli\eV}$).

The crucial difference between QB and PI is the functional course of the oscillation extrema in the 
$\tau_{12}$-$\tau_\text{ref}$-map. For the QB, the extrema run parallel to the 
line $\tau_\text{ref} = \tau_{12}$ (red line), whereas for PI, they follow $\tau_\text{ref} = 2\tau_{12}$ (blue line). 
This property represents a simple way to decide whether our PE signal arises from biexciton or exciton/trion.
However, on the basis of the data shown in Figure~\sref{fig: QB_PI_discussion}{a}, we cannot distinguish between 
QB and PI since the involved decoherence times are to short to observe any oscillations of the PE amplitude. 

The same argument as for the extrema of the PE amplitude holds for the extrema of 
any polarization contrast that exhibits oscillations due to QB or PI. 
Hence, we modelled in Figures~\sref{fig: QB_PI_discussion}{d} and \hyperref[fig: QB_PI_discussion]{(e)} the polarization contrast 
$\rho_2^{\sigma^{\!+}\text{V}}$ from biexciton 
and exciton/trion. 
Figure~\sref{fig: QB_PI_discussion}{d} shows the corresponding 
experimental data for comparison. 
Note, independent of the inhomogeneous broadening, in theory we can observe the polarization contrast for any positive values 
of $\tau_{12}$ and $\tau_\text{ref}$ due to an unlimited dynamic range.
Again, in the case of QB (Figure~\sref{fig: QB_PI_discussion}{d}), 
the extrema run parallel to $\tau_\text{ref} = \tau_{12}$, for PI (Figure~\sref{fig: QB_PI_discussion}{e}) 
parallel to $\tau_\text{ref} = 2\tau_{12}$.
Because of the different damping behaviour of the polarization contrast 
(as discussed in Section~\ref{sec: fitting_biexciton}), 
the experimental data for $\rho_2^{\sigma^{\!+}\text{V}}$ allows us to observe oscillations in the $\tau_{12}$-$\tau_\text{ref}$-map, 
Figure~\sref{fig: QB_PI_discussion}{d}. We can clearly identify that the extrema of the observed 
oscillations run parallel to $\tau_\text{ref} = \tau_{12}$, which rules out the 
possibility of polarization interference. 

\subsection{Modeling for other excitonic complexes}

\begin{landscape}
    \begin{figure}
        \centering 
\includegraphics[scale = 1]{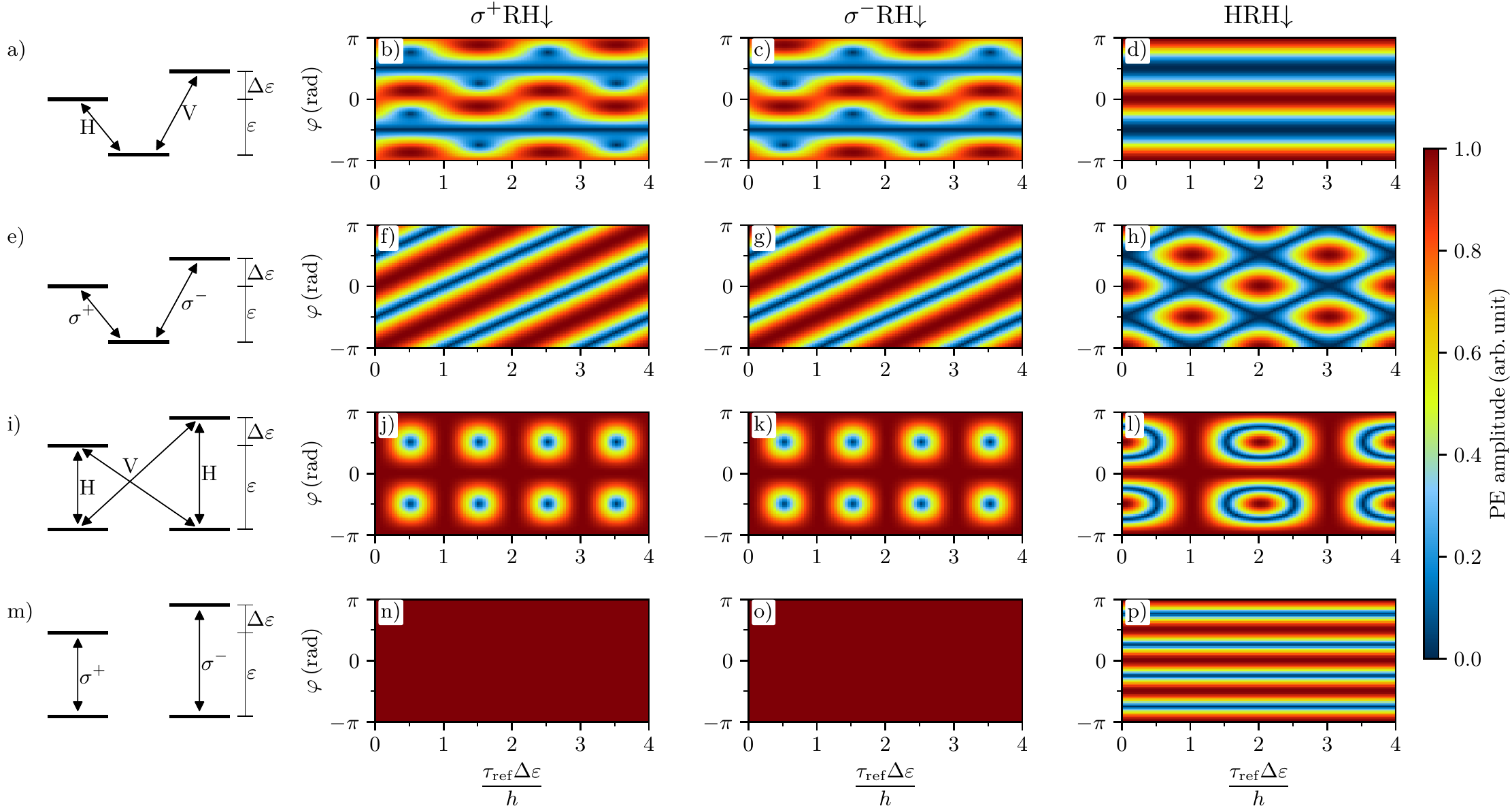}
        \caption{Modeling of expected behaviour of different level schemes in polarimetry
        experiments. The results are arranged in a table. The labels on top of each
        column indicate the polarization configuration. The left column shows the schematics
        of the considered level scheme for each row. Note that we use the labels $\varepsilon$ and 
        $\Delta \varepsilon$ independently for each system.}
        \label{fig: model_exciton_trion}
    \end{figure}    
\end{landscape}

\begin{landscape}
    \begin{figure}
        \centering 
\includegraphics[scale = 1]{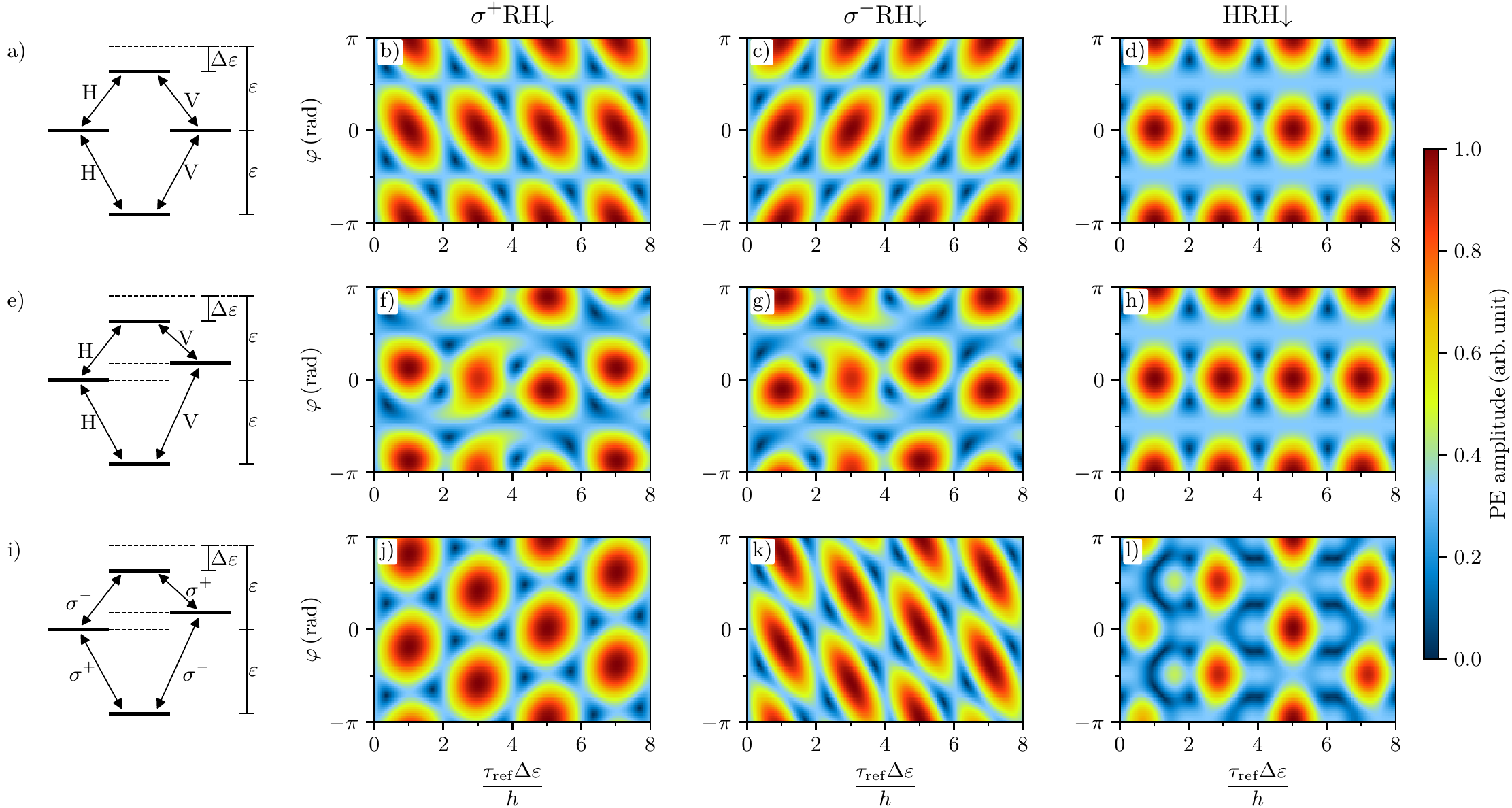}
        \caption{Modeling of expected behaviour of different level schemes in polarimetry
        experiments. The results are arranged in a table. The labels on top of each
        column indicate the polarization configuration. The left column shows the schematics
        of the considered level scheme for each row. Note that we use the labels $\varepsilon$ and 
        $\Delta \varepsilon$ independently for each system.}
        \label{fig: model_biexciton}
    \end{figure}    
\end{landscape}

The outcome of the modeling procedure described in Section~\ref{sec: modeling} is fully determined by 
the Hamiltonian $\mathbf{H}$ and the operator $\mathbf{V}$ describing the interaction with light. 
The combination of both operators can be visualized in terms of level schemes as depicted
in the left columns of Figures~\ref{fig: biexciton_experiment},~\ref{fig: model_exciton_trion}, and~\ref{fig: model_biexciton}.
We applied our modeling procedure to a variety of such level schemes 
that exhibit an energy splitting $\Delta \varepsilon$ and thus give rise to quantum beats. 
We calculated the behaviour of such level schemes in the 
polarization configurations $\sigma^+$RH, $\sigma^-$RH, and HRH, which  
is summarized in the Figures~\ref{fig: model_exciton_trion}
and~\ref{fig: model_biexciton}. Here we can find, that the resulting pictures dramatically change for different schemes. 
These considerations 
let us expect that our experimental method presents a simple and efficient tool 
for a better understanding of unknown systems in terms of energy level arrangements and 
polarization selection rules.

\singlespace
\bibliography{references}

\end{document}